\documentclass[useAMS,onecolumn,usenatbib]{mn2e}
\usepackage{graphicx}
\usepackage{times}
\usepackage{graphicx}
\usepackage{latexsym}
\usepackage{amssymb}

\def\lsim{\lower.5ex\hbox{$\; \buildrel < \over \sim \;$}}
\def\gsim{\lower.5ex\hbox{$\; \buildrel > \over \sim \;$}}
\def\lsim{\lower.5ex\hbox{$\; \buildrel < \over \sim \;$}}
\def\gsim{\lower.5ex\hbox{$\; \buildrel > \over \sim \;$}}
\def\pmb#1{\setbox0=\hbox{$#1$}%
\kern-.025em\copy0\kern-\wd0
\kern.05em\copy0\kern-\wd0
\kern-.025em\raise.0433em\box0}
\def\lsim{\lower.5ex\hbox{$\; \buildrel < \over \sim \;$}}
\def\gsim{\lower.5ex\hbox{$\; \buildrel > \over \sim \;$}}

\def\vc{\bf {{\vert_{(r=r_c)}}}}

\def\egam{$\left\{{\cal E},{\gamma}\right\}$}
\def\eker{$\left[{\cal E},\lambda,\gamma,a\right]$}

\newtheorem{theorem}{Theorem}[section]
\newtheorem{proposition}[theorem]{Proposition}
\newtheorem{conjecture}[theorem]{Conjecture}
\newtheorem{lemma}[theorem]{Lemma}

\title[Hysteresis in shocked black hole accretion]
{Hysteresis effects and diagnostics of the shock formation in low 
angular momentum axisymmetric accretion in the Kerr metric}
\author[Das \& Czerny]
{Tapas K. Das$^{1}$\thanks{tapas@mri.ernet.in}
and Bozena Czerny $^{2}$\thanks{bcz@camk.edu.pl}\\
$^{1}$Harish Chandra Research Institute, Chhatnag Road, Jhunsi,
Allahabad 211019, India\\
$^{2}$Nicolaus Copernicus Astronomical Center, Bartycka 18, 00-716 Warsaw, Poland}
\begin{document}

\maketitle

\label{firstpage}

\begin{abstract}
\noindent
\noindent
The secular evolution of the purely
general relativistic low angular momentum accretion flow around a spinning black hole
is shown to exhibit hysteresis effects.
This confirms that a stationary shock is an 
integral part of such an
accretion disc in the Kerr metric. The equations describing 
the space gradient of the dynamical flow velocity of the accreting 
matter have been shown to be equivalent to a first order autonomous
dynamical systems. Fixed point analysis ensures that such flow must 
be multi-transonic for certain astrophysically relevant initial 
boundary conditions. Contrary to the existing consensus in the 
literature, the critical points and the sonic 
points are proved not to be isomorphic in general, they can form in a
completely different length scales. Physically acceptable global transonic 
solutions must produce odd number of critical points.
Homoclinic orbits for the flow flow possessing multiple 
critical points select
the critical point with the 
higher entropy accretion rate, confirming that the entropy accretion rate 
is the degeneracy removing agent in the system. However, heteroclinic 
orbits are also observed for some special situation, where both the saddle type
critical points of the 
flow configuration possesses identical entropy accretion rate.
Topologies with heteroclinic orbits are thus the only allowed non removable degenerate 
solutions for accretion flow with multiple critical points, 
and are shown to be structurally unstable. 
Depending on suitable initial boundary conditions, a homoclinic trajectory 
can be combined with a standard non homoclinic orbit through an energy 
preserving Rankine-Hugoniot type of stationary shock,
and multi-critical accretion flow then becomes truly 
multi-transonic. An effective 
Lyapunov index has been proposed to analytically confirm why certain class of 
transonic flow can not accommodate shock solutions even if it produces multiple critical 
points.
\end{abstract}


\section{Whither shocked accretion disc?}
\label{sec1}
\noindent
For accretion of matter onto astrophysical black holes, the 
local radial Mach number $M$ of the accreting fluid can be defined as the ratio
of the radial component of the local dynamical flow velocity to that of 
the propagation
of the acoustic perturbation embedded inside the accreting matter.
The flow will be locally subsonic or supersonic according to $M < 1$
or $ >1$. The flow is transonic if at any moment
 it crosses the $M=1$ hypersurface. This happens when a subsonic to 
supersonic or supersonic to
subsonic transition takes place either continuously or discontinuously.
The point(s) where such crossing
takes place continuously is (are) called sonic point(s),
and where such crossing takes place discontinuously are called shocks
or discontinuities.
The particular value
of the radial distance $r$ 
for which {$M=1$}, is referred as the
transonic point or the sonic point, and will be denoted by
$r_s$ hereafter. For $r<r_s$,  infalling matter becomes supersonic. Any
acoustic perturbation created in such a region is destined to be dragged
toward the black hole, and can not escape to the domain 
$r>r_s$. In other words, any co-moving observer from
$r < r_s$ region can not communicate with any observer (co-moving or stationary)
located in the sub-domain
$r>r_s$ by sending any signal which travels with velocity
$v_{\rm signal}{\le}c_s$, where $c_s$ is defined as the velocity of
propagation of the acoustic perturbation (the sound speed) embedded in the
moving fluid. Hence the hypersurface through
$r_s$ is generated by the acoustic null geodesics,
i.e., by the phonon trajectories, and is actually an
acoustic horizon for stationary
configuration,  
which is produced when accreting fluid makes a
transition from subsonic ({$M < 1$}) to the supersonic
({$M > 1$}) state.
At a distance far away from the black hole, accreting material almost
always remains subsonic (except possibly for the supersonic
stellar wind fed accretion) since it possesses negligible dynamical
flow velocity. On the other hand, the flow velocity will approach
the velocity of light $c$ while crossing the event horizon, while the maximum
possible value of sound speed, even for the steepest possible equation
of state, would be $c/\sqrt{3}$, resulting $M>1$ close to the
event horizon.
In order to satisfy such inner boundary condition imposed by the
event horizon, accretion onto black holes
exhibit transonic properties in general, see, e.g., \cite{pri81,skc90,kato-book,fkr02}
for further details.

It will be shown in the subsequent sections that the physical transonic 
accretion solutions can formally be realized as critical solution 
on the phase portrait (spanned by dynamical flow/Mach number and the 
radial distance) of the black hole accretion. It will also be 
discussed, in great details, that usually a critical point is not
identical with a sonic point. 
For sub-Keplerian angular momentum distribution of matter, frequently it 
happens that the critical features are exhibited more than once in the 
phase portrait of a stationary solutions describing an axisymmetric 
black hole accretion. For such situations, the number of critical points, unlike spherical
accretion, may exceed one.
A large pool of literature, dealing with the theory of the black hole accretion 
disc, studied various properties of such flows with multiple critical points
\citep{lt80,az81,boz-pac,boz1,fuk83,fuk87,fuk04,fuk04a,lu85,lu86,bmc86,ak89,abram-chak,
skc90,ky94,yk95,par96,pa97,caditz-tsuruta,das02,bdw04,abd06,sand07}.

For realistic astrophysical systems, such sub-Keplerian
weakly rotating flows
are exhibited in
various physical situations, such as detached binary systems
fed by accretion from OB stellar winds (\cite{ila-shu,liang-nolan}),
semi-detached low-mass non-magnetic binaries (~\cite{bisikalo}),
and super-massive black holes fed
by accretion from slowly rotating central stellar clusters (\cite{ila,ho} and references therein). 
Even for a standard Keplerian
accretion disc, turbulence may produce such low angular momentum flow
(see, e.g.,~\cite{igu},
and references therein).

In supersonic astrophysical flows, perturbation of various kinds may produce
discontinuities. Such discontinuities are said to take place over one or
more surfaces when any dynamical and/or thermodynamic quantity changes
discontinuously as such surfaces are crossed. The corresponding surfaces
are called the surfaces of discontinuity. Certain boundary conditions
are to be satisfied across such surfaces, and according to those conditions,
surfaces of discontinuities are classified into various categories, the
most important being the shock waves or shocks. Such shock waves
are quite often generated in
various kinds of supersonic astrophysical flows having
intrinsic angular momentum, resulting the final state of the flow to be subsonic.
This is because the repulsive centrifugal potential barrier
experienced by such flows is sufficiently strong to brake the infalling
motion and a stationary solution
could be introduced only through a shock. Rotating, transonic
astrophysical fluid
flows are thus believed to be `prone' to the shock formation phenomena.

The non-linear equations describing the steady, inviscid axisymmetric
flow in the Kerr metric can be tailored to form a first order autonomous
dynamical system.
Physical transonic solution in such flows can be represented
mathematically as critical solutions in the velocity (or Mach
number) phase plane of the flow -- they are associated with the critical
points (alternatively known as the fixed points or the equilibrium
points, see \cite{js99} and \cite{diff-eqn-book} for further
details about the fixed point analysis techniques). To maintain the
transonicity such critical points will perforce have to be saddle points,
which will enable a solution to pass through themselves.

Before we proceed further, it is important to introduce a few nomenclatures we will 
be following in this work. When a configuration can have multiple critical points accessible 
to the accretion flow, we will mention it as `multi-critical' accretion. Critical 
points for a non dissipative system can be of two types, either saddle or centre. A saddle 
type critical points allows an accretion flow to pass through it whereas a centre type
critical point does not. Hence the concept of a `sonic point' can only be associated with a 
saddle type critical point and not with a centre type one. No transonic 
solution can pass through a centre type critical point. We will demonstrate that 
a saddle type critical point usually has a sonic point associated with it. As a result,
a multi-critical flow having three critical points, two saddle and one 
centre type, can only have two sonic points. A `multi-transonic' flow, according to our definition,
will be a particular subclass of multi-critical transonic accretion which can pass through two sonic points. 
This is possible, as we will see in subsequent sections, only if a shock forms. Otherwise, 
even if a flow has more than one saddle type critical points, transonic accretion passes 
only through a single sonic point. Such flows will in generally be termed as a `multi-ctritical 
mono-transonic' flow in our work. On the contrary, when the number of the critical point
itself is  one, and that critical point is of saddle type so that a sonic point is also associated with the
flow, we will call it a `mono-critical mono-transonic' flow. We will see in the subsequent sections 
that among the several phases of the transonic accretion flow, a true multi-transonic 
flow forms only for a shocked accretion, otherwise all accretion flows are mono-transonic: 
either of the multi-critical mono-transonic category or of mono-critical mono-transonic 
category. 
All these issues will further be elaborated with greater detail 
in subsequent sections. 

The true multi-transonicity in the Kerr
geometry implies the existence of three critical points and a standing 
shock, and under the
fundamental physical requirement of accretion being a process whereby
a flow solution should connect infinity to the event horizon of the
black hole, the three critical points should be such that there would be
two saddle points flanking a centre type point between themselves.
Originating at a large distance, subsonic accretion
encounters the outermost saddle type sonic point and becomes
supersonic. Subjected to the appropriate perturbative environment, such
supersonic flow encounters a shock and becomes subsonic again. Such
subsonic matter has to pass through another saddle type sonic point
to meet the aforesaid inner boundary condition. For accretion onto
black hole, presence of at least two saddle type {\it sonic} points are
thus a necessary (but not sufficient)
condition for the shock formation.
Multi-transonicity, thus, plays
a crucial role in studying the physics of shock formation and related
phenomena in connection to the black hole accretion processes.

One also
expects that a shock formation in black-hole accretion discs
might be a general phenomenon because shock waves
in rotating astrophysical flows potentially
provide an important and efficient mechanism
for conversion of a significant amount of the
gravitational energy  into
radiation by randomizing the directed infall motion of
the accreting fluid. Hence, the shocks  play an
important role in governing the overall dynamical and
radiative processes taking place in astrophysical fluids and
plasma accreting
onto black holes.

The hot and dense post shock flow is considered to be a powerful
diagnostic tool in understanding various astrophysical phenomena like
the spectral properties of the galactic black hole candidates \citep{ct95}
and that of the supermassive black hole at our Galactic centre 
\citep{monika}, the formation and
dynamics of accretion powered galactic and extra-galactic outflows
\citep{dc99,indra-santa,santa-indra}, shock induced nucleosynthesis
in the black hole accretion disc and the metalicity of the intergalactic
matter (\citep{mukhopadhyay} and references therein), and the origin of the 
quasi-periodic oscillation in galactic sources (\citep{spon-molt,vadawale,okuda1,okuda2}
and references therein).

The study of steady, standing, stationary shock waves produced in black
hole accretion and related phenomena thus acquired an important status
in recent years(~\citep{fuk83,fuk87,fuk04,fuk04a,kat88,c89,ky94,yk95,caditz-tsuruta,
fukumara-suruta,takahashi,das02,lyyy97,lugu,nf89,nagyam08,
nakayama,nagakura,toth}). However, since for the same set of initial boundary conditions describing 
the multi-transonic accretion, there exist flow profiles which may or may not 
accommodate shock solutions, the very existence of accretion flow with shock
is still a matter of debate to the astrophysical community involved in the 
study of the theory of black hole accretion disc. 

This present treatment
{\it confirms}, using a much broader and more comprehensive procedure 
in its scope and 
objectives than those reported previously in the literature, that the 
shock {\it is} an integral part of the black hole axisymmetric accretion
in its most generic form, i.e., for the general relativistic accretion disc 
in the Kerr metric. In the present paper we study the sequence of the stationary 
axisymmetric solutions characterized by the angular momentum 
paying attention to the changes of the flow topology in the extended velocity 
phase space diagram.
By the phrase `extended velocity phase diagram', here we mean the 
two dimensional graphical representation where the radial 
Mach number has been plotted along the ordinate and the radial distance has been 
plotted along the abscissa. The radial Mach number is the ratio 
of the radial velocity measured on the equatorial plane of the accretion disc, and
the height averaged polytropic sound speed defined on the equatorial 
plane of the disc. The radial distance has been measured from the 
black hole along the equatorial plane, and is scaled in the units of $GM_{BH}/c^2$, where $G$ is the 
Universal gravitational constant and $M_{BH}$ is the mass of the black hole
considered, $G$ and $c$ have been scaled to 
be unity in the system of geometrical unit. We discuss in detail the critical solution which divides the family of 
compact transonic flows (solutions with sonic point at the innermost critical point) 
from extended transonic flows (solutions with sonic point  at the outermost 
critical point). We argue that the existence of this critical solution is likely 
to lead to a hysteresis effect when the system chooses the solution with, or 
without a shock, depending on the flow history.
In addition, we have proposed a Lyapunov like treatment, to show that the 
likelihood of the event that a shock will form or not, clearly depends on the 
value of such effective Lyapunov index. 

The plan of the paper is as follows:

In the next section, we will describe the overall flow configuration, and the corresponding 
governing equations will be derived, solution of which, along with a detailed discussion 
of the transonic properties arising out of such solutions,  will be provided in the section \ref{section3}
Section \ref{section4} and \ref{section5} will elaborate the multi-transonic behaviour of such flow. In section 
\ref{section6} we
will describe certain `hysteresis effects' corresponding to the secular evolution of the flow.
Section \ref{section7} will illustrate the details of an Lyapunov like treatment for the state transition in 
transonic accretion solution to establish what prompts a multi-transonic accretion to 
(or not to) contain a standing shock solution. Finally we conclude in the section \ref{section8}.
\section{Dressing up the disc}
\label{sec2}
\subsection{Spacetime geometry and the stress energy tensor}
\label{subsec2.1}
\noindent
To provide a generic description of axisymmetric fluid flow in strong
gravity, one needs to solve the equations of motion for the
fluid and the Einstein equations. As of our present (limited) understanding about the 
detail physics of the black hole accretion flow is concerned, this seems to be 
an enormously difficult, if not impossible, task to accomplish. 
The problem may be made relatively
tractable by assuming the accretion to be non-self
gravitating so that the fluid dynamics may be dealt in a
metric without back-reactions.
The most general form of the energy momentum
tensor for such non self gravitating compressible hydromagnetic astrophysical
fluid (with a frozen in magnetic field) vulnerable to the shear,
bulk viscosity and generalized energy exchange, may be
expressed as \citep{nt73}:
\begin{equation}
{\large\sf T}^{{\mu}{\nu}}={\large\sf T}^{{\mu}{\nu}}_{\sf M}+{\large\sf T}^{{\mu}{\nu}}_{\large\sf B}
\label{eq85}
\end{equation}
where ${\large\sf T}^{{\mu}{\nu}}_{\sf M}$ and ${\large\sf T}^{{\mu}{\nu}}_{\large\sf B}$
are the fluid (matter) part and the Maxwellian
(electromagnetic) part of the energy momentum
tensor.
${\large\sf T}^{{\mu}{\nu}}_{\sf M}$ and ${\large\sf T}^{{\mu}{\nu}}_{\large\sf B}$
may be expressed as:
\begin{equation}
{\large\sf T}^{{\mu}{\nu}}_{\sf M}={\rho}v^{\mu}v^{\nu}+\left(p-\varsigma{\theta}\right)h^{\mu\nu}
-2\eta{\sigma}^{\mu\nu}+{\rm q}^{\mu}v^\nu+v^\mu{{\rm q}^\nu},~
{\large\sf T}^{{\mu}{\nu}}_{\large\sf B}=\frac{1}{8\pi}\left({\rm B}^2v^{\mu}v^\nu+{\rm B}^2h^{\mu\nu}
-2{\rm B}^\mu{\rm B}^\nu\right)
\label{eq86}
\end{equation}
In the above expression, ${\rho}v^{\mu}v^{\nu}$ is the total mass energy density excluding the
frozen-in magnetic field mass energy density as measured in the local rest frame of the
baryons (local orthonormal frame, hereafter LRF,
 in which there is no net baryon flux in any direction).
$ph^{\mu\nu}$ is the anisotropic pressure for incompressible gas (had it been the case that
$\theta$ would be zero). $\varsigma$ and $\eta$ are the co-efficient of bulk viscosity
and of dynamic viscosity, respectively. Hence $-\varsigma{\theta}h^{\mu\nu}$ and
$-2{\eta}{\sigma^{\mu\nu}}$ are the 
bulk viscosity and the viscous shear
stress, respectively. ${\rm q}^{\mu}v^\nu+v^\mu{{\rm q}^\nu}$ is the energy and
momentum flux, respectively, in LRF of the
baryons. In the expression for ${\large\sf T}^{{\mu}{\nu}}_{\large\sf B}$,
${\rm B}^2/8\pi$ in the first term represents the energy density, in the second
term represents the magnetic pressure orthogonal to the magnetic field lines,
and in third term magnetic tension along the field lines (all terms expressed in LRF),
respectively.

Here, the electromagnetic field may be described by the field tensor
${\cal F}^{{\mu}{\nu}}$ and it's dual
${\cal F}^{{\ast}{\mu}{\nu}}$ (obtained from
${\cal F}^{{\mu}{\nu}}$ using Levi-Civita totally antisymmetric tensor
${\epsilon}^{{\mu}{\nu}{\alpha}{\beta}}$)
satisfying the Maxwell equations through the vanishing of the
four-divergence of ${\cal F}^{{\ast}{\mu}{\nu}}$.
A complete description of flow behaviour could thus be obtained
by taking the co-variant derivative of ${\large\sf T}^{{\mu}{\nu}}$
and ${\rho}v^{\mu}$ to obtain the energy momentum
conservation equations and the conservation of baryonic mass.

However, at this stage, the complete solution remains
analytically untenable unless we are compelled
to adopt a number
of simplified approximations.
Our work concentrates on the inviscid
accretion of hydrodynamic fluid.
Hence ${\large\sf T}^{{\mu}{\nu}}$
may be described by
the standard form of the energy momentum (stress-energy)
tensor of a perfect fluid:
\begin{equation}
{\large\sf T}^{{\mu}{\nu}}=\left(\epsilon+p\right)v_{\mu}v_{\nu}+pg_{{\mu}{\nu}},
~
{\rm or,}~{\bf T}=\left(\epsilon+p\right){\bf v}{\otimes}{\bf v}+p{\bf g}
\label{eq87}
\end{equation}
Our calculation will thus be
focused on the stationary
axisymmetric solution of the energy momentum
and baryon number conservation equations
\begin{equation}
{{\large\sf T}^{{\mu}{\nu}}}_{;\nu}
=0;
\;\;\;\;\;
\left({\rho}{v^\mu}\right)_{;\mu}=0,
\label{eq88}
\end{equation}
Specifying the metric to be stationary and axially symmetric,
 the two
generators
 $\xi^{\mu}\equiv (\partial/\partial t)^{\mu}$ and
 $\phi^{\mu}\equiv (\partial/\partial \phi)^{\mu}$ of the temporal and
axial isometry, respectively, are
Killing vectors.

To describe the flow, it is advisable 
to use the Boyer-Lindquist
co-ordinate \cite{boyer},
and an
azimuthally Lorentz boosted orthonormal tetrad basis co-rotating
with the accreting fluid. At this stage we are not interested in non-axisymmetric
disc structure, hence we neglect any gravo-magneto-viscous
non-alignment between the flow angular momentum and the black hole spin angular
momentum. We consider inviscid accretion and denote the constant specific flow angular 
momentum to be $\lambda$. 
Hence, while describing the accretion disc dynamics,
the viscous transport of the angular momentum has not
explicitly been taken into account. Viscosity, however, is quite a subtle
issue in studying the disc accretion.
Even thirty six years after the discovery of
standard accretion disc theory \citep{ss73,nt73},
exact modeling of viscous
transonic black-hole accretion, including
proper heating and cooling mechanisms, is still quite an arduous task, even for a
Newtonian flow, let alone for general relativistic accretion.
Nevertheless, extremely large radial velocity
close to the black hole implies $\tau_{inf}\ll \tau_{visc}$, where $\tau_{inf}$ and
$\tau_{visc}$ are the infall and the viscous time scales, respectively.
Large radial velocities even at larger distances are due to the fact
that the angular momentum content of the accreting fluid
is relatively low \citep{belo,belo1,2003}.
Hence, the assumption of inviscid flow is not unjustified from
an astrophysical point of view. 
However,
one of the most significant effects of the introduction of viscosity
would be the reduction of the angular momentum.
In our work, it has been observed that the location of the sonic points
anti-correlates with $\lambda$, i.e. weakly rotating flow makes the
dynamical velocity gradient steeper, which indicates that for
viscous flow the sonic horizons will be pushed further out and the flow would
become supersonic at a larger distance for the same set of other initial
boundary conditions. Also we found how the shock location (and strength, and other 
related variables) get modified because of the variation of the flow angular 
momentum. Hence in our work, 
we practically have demonstrated the effect of viscosity as well.

We consider the flow to be
`advective', i.e. to possess considerable radial three-velocity.
The above-mentioned advective velocity, which we hereafter denote by $u$
and  consider it to be confined on the equatorial plane, is essentially the
three-velocity component perpendicular to the set of hypersurfaces
$\{\Sigma_v\}$ defined by
$v^2={\rm const}$, where $v$ is the magnitude of the 3-velocity.
Each $\Sigma_v$ is timelike since
its normal $\eta_{\mu}\propto \partial_{\mu} v^2$
is spacelike and may be normalized as
$\eta^{\mu}\eta_{\mu}=1$.

We then define the specific angular momentum $\lambda$ and the angular
velocity $\Omega$ as
\begin{equation}
\lambda=-\frac{v_\phi}{v_t}; \;\;\;\;\;
\Omega=\frac{v^\phi}{v^t}
=-\frac{g_{t\phi}+\lambda{g}_{tt}}{{g_{\phi{\phi}}+\lambda{g}_{t{\phi}}}}\, ,
\label{eq89}
\end{equation}

The metric on the equatorial plane is given by (\cite{nt73})
\begin{equation}
ds^2=g_{{\mu}{\nu}}dx^{\mu}dx^{\nu}=-\frac{r^2{\Delta}}{A}dt^2
+\frac{A}{r^2}\left(d\phi-\omega{dt}\right)^2
+\frac{r^2}{\Delta}dr^2+dz^2 ,
\label{eq90}
\end{equation}
where $\Delta=r^2-2r+a^2, ~A=r^4+r^2a^2+2ra^2$,
and $\omega=2ar/A$, $a$ being the Kerr parameter related to the black-hole spin.
The normalization condition $v^\mu{v}_\mu=-1$, together with
the expressions for
$\lambda$ and $\Omega$  in (\ref{eq89}), provides the relationship between the
advective velocity $u$ and the temporal component of the four velocity
\begin{equation}
v_t=
\left[\frac{Ar^2\Delta}
{\left(1-u^2\right)\left\{A^2-4\lambda arA
+\lambda^2r^2\left(4a^2-r^2\Delta\right)\right\}}\right]^{1/2} .
\label{eq91}
\end{equation}
\noindent
In order to solve (\ref{eq88}), we need to specify a realistic equation of
state. In this work, we concentrate on polytropic accretion, and some 
relevant details of the flow thermodynamics is presented in the next section.

It is to be noted here that the use of the polytropic equation of state 
in describing the thermodynamic properties of the accreting matter is
quite common in the theory of relativistic black hole accretion.
However, one also understands that the polytropic
accretion is not the only choice to describe such
accretion. Equations of state other than the adiabatic one,  such as
the isothermal equation ~\citep{yk95}
or the two-temperature plasma ~\citep{manmoto},
have also been used to for this purpose.
\subsection{The thermodynamics of the flow}
\label{subsection2.2}
\noindent
The polytropic equation is taken to be of the following form
\begin{equation}
p=K{\rho}^\gamma ,
\label{eq92}
\end{equation}
where the polytropic index $\gamma$ (equal to the ratio of the two specific
heats $c_p$ and $c_v$) of the accreting material is assumed to be constant throughout the fluid.
A more realistic model of the flow 
would perhaps require  a variable polytropic index having a 
functional dependence on the radial
distance, i.e. of the functional form $\gamma{\equiv}\gamma(r)$; see, e.g., 
\citep{ryu-indra} and references therein for further discussions. However, we  have performed the
calculations for a sufficiently large range of $\gamma$ and we believe
that all astrophysically relevant
polytropic indices are covered in our analysis.

The constant $K$ in (\ref{eq92}) may be 
related to the specific entropy of the fluid,
 provided there is no entropy 
generation in the flow. 
If in addition to (\ref{eq92}) the
Clapeyron equation for an ideal gas 
holds
\begin{equation}
p=\frac{\kappa_B}{{\mu}m_p}{\rho}T\, ,
\label{eq93}
\end{equation}
 where $T$ is the locally measured temperature, $\mu$  the mean molecular weight,
$m_H{\sim}m_p$  the mass of the hydrogen atom, then the specific entropy, i.e. the entropy 
per particle, is given by ~\citep{landau}:
\begin{equation}
\sigma=\frac{1}{\gamma -1}\log K+
\frac{\gamma}{\gamma-1}+{\rm constant} ,
\label{eq94}
\end{equation}
where the constant depends on the chemical composition of the 
accreting material. 
Equation (\ref{eq94}) confirms that $K$ in (\ref{eq92})
is  a measure of the specific entropy of the accreting matter.

The specific enthalpy of the accreting matter can now be defined as
\begin{equation}
h=\frac{\left(p+\epsilon\right)}{\rho}\, ,
\label{eq95}
\end{equation}
where the energy density $\epsilon$ includes the rest-mass density and the internal 
energy and may be written as
\begin{equation}
\epsilon=\rho+\frac{p}{\gamma-1}\, .
\label{eq96}
\end{equation}
The adiabatic speed of sound is defined by
\begin{equation}
c_s^2=\frac{{\partial}p}{{\partial}{\epsilon}}{\Bigg{\vert}}_{\rm 
constant~entropy}\, .
\label{eq97}
\end{equation}
From (\ref{eq96}) we obtain
\begin{equation}
\frac{\partial{\rho}}{\partial{\epsilon}}
=\left(
\frac{\gamma-1-c_s^2}{\gamma-1}\right) .
\label{eq98}
\end{equation}
Combination of (\ref{eq97}) and (\ref{eq92}) gives
\begin{equation}
c_s^2=K{\rho}^{\gamma-1}{\gamma}\frac{\partial{\rho}}{\partial{\epsilon}}\, ,
\label{eq99}
\end{equation}
Using the above relations, one obtains the expression for the specific enthalpy 
\begin{equation}
h=\frac{\gamma-1}{\gamma-1-c_s^2}\, .
\label{eq100}
\end{equation}
The rest-mass density $\rho$, the pressure $p$, the temperature $T$
of the flow and the energy density $\epsilon$ 
may be expressed in terms of the speed of sound  $c_s$ as
\begin{equation}
\rho=K^{-\frac{1}{\gamma-1}}
\left(\frac{\gamma-1}{\gamma}\right)^{\frac{1}{\gamma-1}}
\left(\frac{c_s^2}{\gamma-1-c_s^2}\right)^{\frac{1}{\gamma-1}},
\label{eq101}
\end{equation}
\begin{equation}
p=K^{-\frac{1}{\gamma-1}}
\left(\frac{\gamma-1}{\gamma}\right)^{\frac{\gamma}{\gamma-1}}
\left(\frac{c_s^2}{\gamma-1-c_s^2}\right)^{\frac{\gamma}{\gamma-1}},
\label{eq102}
\end{equation}
\begin{equation}
T=\frac{\kappa_B}{{\mu}m_p}
\left(\frac{\gamma-1}{\gamma}\right)
\left(\frac{c_s^2}{\gamma-1-c_s^2}\right),
\label{eq103}
\end{equation}
\begin{equation}
\epsilon=
K^{-\frac{1}{\gamma-1}}
\left(\frac{\gamma-1}{\gamma}\right)^{\frac{1}{\gamma-1}}
\left(\frac{c_s^2}{\gamma-1-c_s^2}\right)^{\frac{1}{\gamma-1}}
\left[
1+\frac{1}{\gamma}
\left(
\frac{c_s^2}{\gamma-1-c_s^2}
\right)
\right].
\label{eq104}
\end{equation}

We now need to define a specific geometrical structure of the disc, which 
has been done in the next section.
\subsection{The disc geometry}
\label{subsection2.3}
\noindent
We assume that
the disc has a radius-dependent local
thickness $H(r)$, and its central plane coincides with
the equatorial plane of the black hole.
It is a standard practice
in accretion disc theory
(\cite{mkfo84,pac87,acls88,ct93,ky94,abn96,nkh97,wiita98,hawley-krolik,
armitage} and many others)
to
use the vertically averaged
model in
describing the black-hole accretion discs where the equations of motion
apply to the equatorial plane of the black hole.
We follow the same
procedure here.
The thermodynamic
flow variables are averaged over the disc height,
i.e.,
a thermodynamic
quantity $y$ used in our model is vertically averaged over the disc height
as\\
$$
\bar{y}=\frac{\int^H(r)_0 (ydh)}{\int^H(r)_0 H(r)}.
$$
We follow  \cite{discheight}
to derive an expression for the disc height $H(r)$
in our flow geometry since the relevant equations in
\cite{discheight}
are non-singular on the horizon and can accommodate both the axial and
a quasi-spherical flow geometry. In the Newtonian framework, the disc height
 in vertical
equilibrium is obtained from the $z$ component of the non-relativistic Euler
equation where all the terms involving velocities and the
higher powers of $\left({z}/{r}\right)$ are neglected.
In the case of a general relativistic disc, the vertical pressure
gradient in the comoving frame is compensated by the tidal gravitational
field. We then obtain the disc height
\begin{equation}
H(r)=\sqrt{\frac{2}{\gamma + 1}} r^{2} \left[ \frac{(\gamma - 1)c^{2}_{s}}
{\{\gamma - (1+c^{2}_{s})\} \{ \lambda^{2}v_t^2-a^{2}(v_{t}-1) \}}\right] ^{\frac{1}{2}} ,
\label{eq105}
\end{equation}
which, by making use of
(\ref{eq91}),
may be be expressed in terms of
the advective velocity $u$.
\section{Governing equations and the solution procedure}
\label{section3}
\subsection{The first integrals of motion}
\label{subsection3.1}
\noindent
From analytical perspective, problems in black hole accretion fall
under the general class of nonlinear dynamics 
\cite{rb02,ap03,ray03a,ray03b,rbcqg05a,rbcqg05b,crd06,rbcqg06,rbcqg07a,br07,gkrd07,jkb09},
since accretion describes the dynamics
of a compressible astrophysical fluid, governed by a set of
nonlinear differential equations. Physical transonic solutions can
be mathematically realized as critical solutions in the phase portrait
of the flow. To obtain such critical solutions, it is first convenient
to construct a set of first integrals of motion.

The temporal  component of the energy momentum tensor conservation equation
leads to the
constancy of the total specific energy of the accretion flow along each streamline of the flow.
This  specific energy is denoted by 
${\cal E}$ (which actually is the relativistic analogue of 
the Bernoulli's constant, and can be expressed in terms of the flow 
enthalpy and $v_t$, as ${\cal E}=hv_t$, see \cite{anderson} 
for further detail), and hence
from (\ref{eq91}) and (\ref{eq100}) it follows that:
\begin{equation}
{\cal E} =
\left[ \frac{(\gamma -1)}{\gamma -(1+c^{2}_{s})} \right]
\sqrt{\left(\frac{1}{1-u^{2}}\right)
\left[ \frac{Ar^{2}\Delta}{A^{2}-4\lambda arA +
\lambda^{2}r^{2}(4a^{2}-r^{2}\Delta)} \right] } \, .
\label{eq107}
\end{equation}
The rest-mass accretion rate ${\dot M}$ is obtained by integrating the relativistic
continuity equation. 
One finds
\begin{equation}
{\dot M}=4{\pi}{\Delta}^{\frac{1}{2}}H{\rho}\frac{u}{\sqrt{1-u^2}} \, ,
\label{eq108}
\end{equation}
Here, we adopt the sign convention that a positive $u$ corresponds to
accretion.
The entropy accretion rate ${\dot \Xi}$
can be expressed as:
\begin{equation}
{\dot \Xi}
 = \left( \frac{1}{\gamma} \right)^{\left( \frac{1}{\gamma-1} \right)}
4\pi \Delta^{\frac{1}{2}} c_{s}^{\left( \frac{2}{\gamma - 1}\right) } \frac{u}{\sqrt{1-u^2}}\left[\frac{(\gamma -1)}{\gamma -(1+c^{2}_{s})}
\right] ^{\left( \frac{1}{\gamma -1} \right) } H(r)
\label{eq109}
\end{equation}
One can solve the conservation equations for ${\cal E}, {\dot M}$ and
${\dot \Xi}$ to obtain the complete accretion profile.

We thus have two primary first integrals of motion along the 
streamline -- the specific energy of the flow ${\cal E}$ and the 
mass accretion rate ${\dot M}$. Even in the absence of creation 
or annihilation of matter, the entropy accretion rate ${\dot \Xi}$ 
is not a generic first integral of motion. As the expression for 
${\dot \Xi}$
contains the quantity $K{\equiv}p/{\rho}^{\gamma}$, which is a measure
of the specific entropy of the flow, the entropy accretion rate ${\dot \Xi}$
remains constant throughout the flow only if the entropy per particle remains 
locally invariant. This condition may be violated if the accretion is 
accompanied by a shock. Thus ${\dot \Xi}$
is conserved for shock free polytropic accretion (and wind) and 
becomes discontinuous (actually, increases) at the shock location,
if such a shock is formed. However, ${\dot \Xi}$
is of utmost importance for our purpose, since it acts as the 
degeneracy remover for the multi-critical accretion, 
see subsequent discussions for further detail.

Being equipped with the disc geometry and the integrals of motion, we will
now discuss the transonicity of the flow in the next section.
\subsection{Velocity gradient and the transonicity}
\label {subsection3.2}
\noindent
The gradient of the acoustic velocity can be computed by
differentiating (\ref{eq109}) and can be obtained as:
\begin{equation}
\frac{dc_s}{dr}=
\frac{c_s\left(\gamma-1-c_s^2\right)}{1+\gamma}
\left[
\frac{\chi{\psi_a}}{4} -\frac{2}{r}
-\frac{1}{2u}\left(\frac{2+u{\psi_a}}{1-u^2}\right)\frac{du}{dr} \right]
\label{eq110a}
\end{equation}
The dynamical velocity gradient can then be calculated by differentiating (\ref{eq107})
with the help of (\ref{eq110a}) as:
\begin{equation}
\frac{du}{dr}=
\frac{\displaystyle
\frac{2c_{s}^2}{\left(\gamma+1\right)}
  \left[ \frac{r-1}{\Delta} + \frac{2}{r} -
         \frac{v_{t}\sigma \chi}{4\psi}
  \right] -
  \frac{\chi}{2}}
{ \displaystyle{\frac{u}{\left(1-u^2\right)} -
  \frac{2c_{s}^2}{ \left(\gamma+1\right) \left(1-u^2\right) u }
   \left[ 1-\frac{u^2v_{t}\sigma}{2\psi} \right] }},
\label{eq110}
\end{equation}
\noindent where (see \citep{bdw04} for further detail):
\begin{eqnarray}
\psi=\lambda^2{v_t^2}-a^2\left(v_t-1\right),~
\psi_a=\left(1-\frac{a^2}{\psi}\right),~
\sigma = 2\lambda^2v_{t}-a^2,
~
& & \nonumber \\
\chi =
\frac{1}{\Delta} \frac{d\Delta}{dr} +
\frac{\lambda}{\left(1-\Omega \lambda\right)} \frac{d\Omega}{dr} -
\frac{\displaystyle{\left( \frac{dg_{\phi \phi}}{dr} + \lambda \frac{dg_{t\phi}}{dr} \right)}}
     {\left( g_{\phi \phi} + \lambda g_{t\phi} \right)}.
\label{eq111}
\end{eqnarray}
\noindent
A real physical transonic flow must be smooth everywhere, except possibly at a
shock. Hence, if the denominator of the equation (\ref{eq110}) vanishes at a 
point (for certain value of $r$), the numerator must also vanish at that 
point to ensure the physical continuity of the flow. One thus arrives at a
critical point (or the fixed point/equilibrium point -- by borrowing the terminology 
used in the theory of dynamical systems) conditions by simultaneously making the numerator and
the denominator equal to zero. 
The critical point conditions can thus be obtained as:
\begin{equation}
{c_{s}}_{\vc}={\left[\frac{u^2\left(\gamma+1\right)\psi}
                                  {2\psi-u^2v_t\sigma}
                       \right]^{1/2}_{\vc}  },
~~u{\vc}= {\left[\frac{\chi\Delta r} {2r\left(r-1\right)+ 4\Delta} \right]
^{1/2}_{\rm r=r_c}  },
\label{eq112}
\end{equation}
\noindent
For any value of
\eker,
substitution of the values of $u{\vc}$ and $c_{s}{\vert}_{\rm r=r_c}$  in terms of $r_c$
in the expression
for ${\cal E}$ (\ref{eq107}),
provides
a polynomial in $r_c$, the solution of which determines
the location of the critical point(s) $r_c$.
\subsection{Non isomorphism of the critical and the sonic points}
\label{subsection3.3}
\noindent
It is obvious from (\ref{eq112}) that
$u_{\vc}{\ne}{c_s}_{\rm r=r_c}$, and hence the Mach number at the critical point is {\it not}
equal to unity in general. This phenomena can more explicitly be
demonstrated for relativistic disc accretion onto a Schwarzschild black hole.
For Schwarzschild metric,
we
calculate the Mach number of the flow at the critical point as
\begin{equation}
M_c=
\sqrt{
\left({\frac{2}{\gamma+1}}\right)
\frac
{{f_{1}}(r_c,\lambda)}
{{{f_{1}}(r_c,\lambda)}+{{f_{2}}(r_c,\lambda)}}
}\, .
\label{eq115}
\end{equation}
where
\begin{equation}
{f_{1}}(r_c,\lambda) = \frac{3r_c^{3}-2\lambda^{2}r_c+
3{\lambda^2}}{r_c^{4}-\lambda^{2}r_c(r_c-2)},~
{f_{2} } (r_c,\lambda) = \frac{2r_c-3}{r_c(r_c-2)} -
\frac{2r_c^{3}-\lambda^{2}r_c+\lambda^{2}}{r_c^{4}-\lambda^{2}r_c(r_c-2)}
\label{eq115a}
\end{equation}
Clearly, $M_c$ is generally not equal to unity, and for $\gamma\geq 1$, is always less
than one, and it can easily be shown that for a realistic choice of 
initial boundary conditions, the separation between the critical and the sonic points
can be as high as several hundred gravitational radii. .

Hence, the sonic points are {\it not} isomorphic to the critical points in general,
neither numerically, nor topologically, and  we categorically distinguish 
a sonic point from a critical point.
In the literature on transonic black-hole accretion discs, the concepts of critical
and sonic points are often made identical by erroneously defining an `effective' sound speed
leading to the `effective'  Mach number (for further details, see, eg.
\cite{mkfo84,c89}) for an `effective' geometry.
For realistic flow in general relativity, the exact sonic condition 
should be obtained for at the location where the bulk flow velocity
has to match the local unscaled speed of propagation of the acoustic 
perturbation. 
We thus strongly disagree to accept the synonymous use of the terms 
critical and sonic points for the following reasons: 

In the existing literature on non general relativistic pseudo-Schwarzschild
transonic disc accretion,
the Mach number at the critical point  turns out to be a function of
$\gamma$ only, and hence $M_c$ remains  constant  if $\gamma$ is constant.
For example,
use of the \cite{pw80} pseudo-Schwarzschild potential to
describe the adiabatic accretion phenomena leads to
(see \cite{das02}, and references therein for further details of such
calculation)
\begin{equation}
M_c=\sqrt{\frac{2}{\gamma+1}}\, .
\label{eq116}
\end{equation}
The above expression does not depend on the location of the
critical point and depends only on the value of the
adiabatic index chosen to describe the flow. Note that
for isothermal accretion $\gamma=1$, and hence the sonic points and
the critical points are identical.

However, the quantity $M_c$ in Eq. (\ref{eq115})
as well as when calculated for (\ref{eq112}),
is clearly a function of $r_c$, and hence, generally, it takes  different
values for different $r_c$ for transonic accretion.
The Mach numbers at the critical (saddle type) points are 
thus {\it functional}, and not functions,  
of the initial boundary conditions.
The difference between the
radii of the critical
point and the sonic point may be quite significant.
One defines the
radial difference of the critical and the sonic point
(where the Mach number is exactly equal to unity) as
\begin{equation}
{\Delta}r_c^s=|r_s-r_c|.
\label{eq117}
\end{equation}
The quantity ${\Delta}r_c^s$ may be
 a complicated  function of \eker, the  form of which can not
be expressed analytically, but it can easily be evaluated 
using numerical integrations, and we have done so. We estimate the 
${\Delta}r_c^s$ can be as large as $10^2$ $r_g$,
or even more,
for flow passing through the outer critical/sonic points.

The radius $r_s$ in Eq. (\ref{eq117}) is the radius of the
sonic point  corresponding to the same \eker for which the
radius of the critical point $r_c$ is evaluated.
 Note, however, that since $r_s$ is calculated by integrating the
flow from $r_c$ upto the point where the Mach number exactly becomes unity, 
${\Delta}r_c^s$ is defined only for saddle-type
critical points.
A physically acceptable transonic solution
can be constructed only through a saddle-type critical point,
and not through a centre type critical point.
Hence the concept of a sonic point corresponding to a centre 
type critical point is meaningless, and multi-transonic accretion possesses
{\it two} sonic points (and three critical points). 

\begin{lemma}
\noindent
Physically acceptable global transonic
solutions must always produce odd number of critical points,
whereas even number of sonic points are produced for the multi-transonic
accretion. Odd number of sonic point is observed for mono-transonic accretion,
however, this is a trivial case because the number of sonic point is unity.
\end{lemma}

However, for saddle-type
critical points, $r_c$ and $r_s$ should always have one-to-one correspondence,
in the sense that
every  critical point that  allows a steady solution to pass through it
is accompanied by a sonic point, generally located at a
different radial distance.

It is worth emphasizing that the distinction between critical and
sonic points is a direct manifestation of the non-trivial
functional dependence of the disc thickness on
the fluid velocity, the sound speed
and the radial distance, i.e., on the disc geometry as well 
as the equation of state in general.
In the simplest idealized case when
the disc thickness is assumed to be constant,
or the infalling matter being described using the isothermal
equation of state,
one would expect no distinction
between critical and sonic points.
In this case, as
has  been demonstrated for a thin disc
accretion onto the Kerr black hole  \cite{abd06},
the quantity $\Delta r_c^s$ vanishes identically for any astrophysically
relevant value of \eker.
\begin{figure}
\vskip -5.0truecm
\begin{center}
\includegraphics[scale=0.7, angle=0]{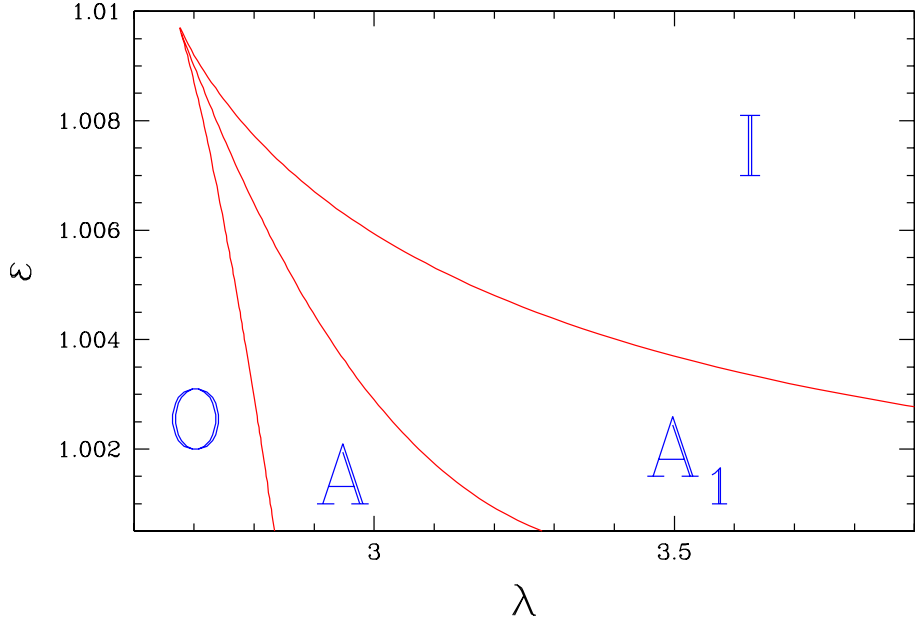}
\caption{\label{f1} \small{
The two dimensional parameter space division for low angular momentum 
axisymmetric accretion in the Kerr metric. The parameter space is spanned by the conserved
specific energy of the flow ${\cal E}$ and the specific angular momentum $\lambda$, and has 
been drawn for a fixed value of $\gamma=4/3$ and $a=0.3$. The {\bf O} and the {\bf I} 
regions represents the mono-critical mono-transonic accretion solutions 
passing through the saddle type outer and the inner critical (sonic) points, respectively. 
The region marked by {\bf A} represents multi-critical accretion, and can further 
be divided (not shown in the figure) into two different regions, ${\bf A_{\sf NS}}$ where 
shock does not form even if the flow has three critical points, and ${\bf A_{\sf S}}$, which 
accommodates {\it true} multi-transonic accretion with general relativistic 
Rankine Hugoniot shock. The region marked by ${\bf A_1}$ represents mono-transonic 
accretion passing through the inner sonic point and accompanied by an additional pair 
of non-accessible critical points and a homoclinic orbit through one of those, see
text for further details.
}}
\end{center}
\end{figure}
\section{March toward the multi-transonicity}
\label{section4}
\subsection{The parameter space diagram}
\label{subsection4.1}
\noindent
By substituting the value of the bulk velocity and the acoustic velocity evaluated 
at the critical points to the expression for the energy first integral, one obtains 
a quasi polynomial (not an exact polynomial in strict mathematical sense, since
it contains fractional power terms) in $r_c$ parametrized by \eker. Solution of 
such a quasi polynomial (with
additional physical constraint that all solutions must be real, positive and
will be greater than $r_g$), provides the critical points for
general relativistic axisymmetric accretion in the Kerr metric. One finds at 
most three critical points for relativistic disc accretion for some
values of \eker. 

The above mentioned quasi polynomial, and hence the critical points, are completely 
determined by four parameters \eker. A four dimensional parameter 
hyper-space is thus required to be looked upon to obtain the global set of 
transonic solutions. For the shake of convenience, we analyze a two dimensional 
projection of such four dimensional hyperspace. Since ${\cal E},\lambda,\gamma,a$ are 
mutually orthogonal (i.e., choice of one among those four parameters does not influence 
the choice of the other parameter(s)), a total number of $^4C_2$ different choices 
are allowed for selecting such projections, each characterized with two different 
parameters out of the four parameter set \eker, by keeping the other two parameters at 
a pre defined fixed value. In this work, we choose to project the hyperspace on 
an $\left[{\cal E}-\lambda\right]$ plane by keeping $\gamma=4/3$ and $a=0.3$. 
However, similar projected $\left[{\cal E}-\lambda\right]$ submanifold can 
routinely
be analyzed for other values of $\gamma$ and $a$.  

To begin with, we first set the astrophysically relevant 
bounds on {\egam} to model the realistic situations
encountered in astrophysics.
Since the specific energy ${\cal E}$  includes the rest-mass energy,
${\cal E}=1$ is the lower bound which corresponds to a flow with zero
thermal energy at infinity.
Hence, the values ${\cal E}<1$, corresponding to the negative energy accretion states, would be allowed if a mechanism
for a radiative extraction of the rest-mass energy existed.
The possibility of such an extraction
would in turn imply the presence of the 
viscosity or other dissipative mechanisms in the fluid.
Since we concentrate only on non-dissipative flows,
we exclude ${\cal E}<1$.
On the other hand, although almost all ${\cal E}>1$ are theoretically allowed,
 large values of ${\cal E}$ represent flows starting from infinity
with very high thermal energy.
In particular, ${\cal E}>2$ accretion represents enormously
hot flow configurations at very large distance from the black hole,
which are not properly conceivable in realistic astrophysical situations.
Hence, we set $1{\lsim}{\cal E}{\lsim}2$.

The physical lower bound on the polytropic index is
$\gamma=1$,  which  corresponds to isothermal accretion
where accreting fluid remains optically thin.
Hence, the values $\gamma<1$ are not realistic in accretion
astrophysics.
On the other hand,
$\gamma>2$ is possible only for superdense matter
with a very large magnetic
field
and a direction-dependent anisotropic pressure.
The presence of a magnetic field would in turn
require solving the
 general relativistic
magneto-hydrodynamic
equations,  which
is beyond the scope of this paper.
Thus, we set
$1{\lsim}\gamma{\lsim}2$.  However,
astrophysically
preferred
values of $\gamma$ for realistic black-hole accretion range from
$4/3$ (ultra-relativistic)
to $5/3$ (purely non-relativistic flow) \citep{fkr02}.
 Hence, we mainly focus on the parameter range
\begin{equation}
\left[1{\lsim}{\cal E}{\lsim}2,~\frac{4}{3}{\le}\gamma{\le}\frac{5}{3}\right].
\label{egam}
\end{equation}

In figure 1, the regions marked by 
O and I correspond to the 
formation of a single critical point, and hence the {\it mono-transonic}
disc accretion is produced for such region. 
In  the region marked by
{\bf I}, the critical points are called `inner type' critical points since
these points are  formed sufficiently close to the event horizon,
in most of the cases even closer than the innermost stable circular orbit 
(ISCO).
In the region marked by {\bf O}, the
critical points are called `outer type' critical points, because these points are
located relatively far from the black hole.
Depending on the value of
\eker, an outer critical point may be as far as $10^6r_g$,
or more. 

The outer type critical points for the mono-transonic region are formed, 
as is obvious from the figure, for sufficiently weakly-rotating flows. For 
sufficiently low angular 
momentum, accretion flow contains less amount of rotational energy, thus 
most of the kinetic energy in utilized to increase the radial 
dynamical velocity $u$ at a faster rate, leading to a higher value 
of $du/dr$. Under such circumstances, the dynamical velocity $u$ becomes
large enough to overcome the acoustic velocity $c_s$ at a larger radial distance
from the event horizon, leading to the generation of supersonic flow at a large
value of $r$, which results the formation of the sonic point 
(and hence the corresponding critical point) far away from the black hole event horizon.
On the contrary, the inner type critical points are formed,
as is observed from the figure, for strongly rotating flow in general. Owing to 
the fact that such flow would posses a large amount of rotational energy, only a small 
fraction of the total specific energy of the flow will be spent to increase the 
radial dynamical velocity $u$. Hence for such flow, $u$ can overcome $c_s$ only at a 
very small distance (very close to the event horizon) where the intensity of the 
gravitational field becomes enormously large, producing a very high value of the 
linear kinetic energy of the flow (high $u$), over shedding the contribution to the 
total specific energy from all other sources. However, from the figure it is 
also observed that the inner type sonic points are formed also for moderately 
low values of the angular momentum as well (especially in the region close to 
the vertex of the wedge shaped zone marked by ${\sf A_1}$). For such regions, the total conserved specific 
energy is quite high. In the asymptotic limit, the expression for the total specific 
energy is governed by the Newtonian construct, and one can have:
\begin{equation}
{\cal E}=
\left(\frac{u^2}{2}\right)_{\rm linear}
+
\left(\frac{c_s^2}{\gamma-1}\right)_{\rm thermal}
+
\left(\frac{\lambda^2}{2r^2}\right)_{\rm rotational}
+
\left(\Phi\right)_{\rm gravitational}
\label{eqN1}
\end{equation}
where $\Phi$ is the gravitational potential energy in the 
asymptotic limit.
From (\ref{eqN1}) it is obvious that at a considerably large 
distance from the black hole, the contribution to the total energy of 
the flow comes mainly (rather entirely) from the thermal energy. 
A high value of ${\cal E}$ (flow energy in excess to its rest mass energy)
corresponds to a `hot' flow starting from infinity. Hence the acoustic velocity
corresponding to the `hot' flow obeying such outer boundary condition would 
be quite large. For such accretion, flow has to travel a large distance 
subsonically and can acquire a supersonic dynamical velocity 
$u$ only at a very close proximity to the event horizon, where the gravitational 
pull would be enormously strong.

The $\left[{\cal E},\lambda\right]$ corresponding to the  
wedge shaped regions marked by A and ${\sf A_1}$ produces three critical points, among which the
largest and the smallest values correspond to the saddle type, the outer
$r_c^{out}$ and the inner $r_c^{in}$, critical points respectively. The 
centre type middle critical point, $r_c^{mid}$, which is unphysical 
in the sense that no steady transonic solution passes through it, lies 
in between $r_c^{in}$ and $r_c^{out}$. 

For the kind of accretion flow considered in this work, the critical points of
the phase trajectories can be identified first, following which a linearized 
study in the neighbourhood of these critical points may be carried out, to 
develop a complete and rigorous mathematical classification scheme to identify
whether a critical point is of saddle type or of centre type. 
Global understanding of the flow topologies will then necessitate a full 
numerical integration of the non-linear equations of the flow, which has 
successfully been performed in this work. For the shake of completeness
and to have a better understanding of the 
multi-transonic behaviour, it will not be unjustified to briefly
discuss the classification scheme for the various kind of critical points which may 
be produced in relativistic accretion disc around a Kerr black hole. Following 
\cite{gkrd07}, the methodology for performing such classification is 
presented below. 
\subsection{Classification scheme for the critical points}
\label{subsection4.2}
\noindent
Eq. (\ref{eq110}) could be reformulated as 
\begin{equation}
\frac{du^{2}}{dr}={\frac{\frac{2}{\gamma
+1}c_{s}^{2}\left[\frac{g^{\prime}_{1}}{g_{1}}-\frac{1}{g_{2}}\frac{\partial
g_{2}}{\partial
r}\right]-\frac{f^{\prime}}{f}}
{\frac{1}{1-u^{2}}\left(1-\frac{2}{\gamma
+1}\frac{c_{s}^{2}}{u^{2}}\right)+\frac{2}{\gamma
+1}\frac{c_{s}^{2}}{g_{2}}\left(\frac{\partial g_{2}}{\partial
u^{2}}\right)}}
\label{mod1}
\end{equation}
\begin{equation}
\frac{du^{2}}{d{\bar{\tau}}}=\frac{2}{\gamma
+1}c_{s}^{2}\left[\frac{g^{\prime}_{1}}{g_{1}}-\frac{1}{g_{2}}\frac{\partial
g_{2}}{\partial
r}\right]-\frac{f^{\prime}}{f}
\label{mod2}
\end{equation}
with the primes representing the total derivatives with respect to $r$,
and $\bar{\tau}$ is an arbitrary mathematical parameter.
Here, 
\begin{eqnarray}
f(r)=\frac{Ar^{2}\Delta}{A^{2} - 4\lambda arA +
\lambda^{2}r^{2}(4a^{2}-r^{2}\Delta)},~ && \nonumber \\
g_{1}(r)=\Delta r^{4},~g_{2}(r,u)=\frac{\lambda^{2}f}{1-u^{2}}
-\frac{a^{2}f^{\frac{1}{2}}}{\sqrt{1-u^{2}}} + a^{2}
\label{mod3}
\end{eqnarray}

The critical conditions are obtained with the simultaneous vanishing
of the right hand side, and the coefficient of ${d(u^2)/dr}$ in the left 
hand side in (\ref{mod1}). This will provide
\begin{equation}
\left|
\frac{2c_s^2}{\gamma+1}
\left[\frac{g_1^{\prime}}{g_1}-\frac{1}{g_2}\left(\frac{\partial{g_2}}{\partial{r}}\right)\right]
-\frac{f^{\prime}}{f}
\right|_{\rm r=r_c} 
=
\left|
\frac{1}{1-u^2}\left(1-\frac{2}{\gamma+1}\frac{c_s^2}{u^2}\right)
+\frac{2}{\gamma+1}\frac{c_s^2}{g_2}\left(\frac{\partial{g_2}}{\partial{u^2}}\right)
\right|_{\rm r=r_c}=0
\label{mod4}
\end{equation}
as the two critical point conditions.
Some simple algebraic manipulations shows that
\begin{equation}
u_c^2=\frac{f^{\prime}g_1}{f{g_1^{\prime}}},
\label{mod5}
\end{equation}
following which $c_{s}^2|_{\rm r=r_c}$ can be rendered as a function of $r_c$ only, 
and further, by use of (\ref{eq107})., $r_c$, $c_{sc}^2$ and $u_c^2$ can 
all be fixed in terms of the constants of motion like ${\cal E}$, $\gamma$, 
$\lambda$ and $a$. Having fixed the critical points it should now be 
necessary to study their nature in their phase portrait of $u^2$ 
versus $r$. To that end one applies a perturbation about the fixed point 
values, going as,
\begin{equation}
u^2=u^2|_{\rm r=r_c}+\delta{u^2},~c_s^2=c_s^2|_{\rm r=r_c}+\delta{c_s^2},~
r=r_c+\delta{r}
\label{mod6}
\end{equation}
in the parametrized set of autonomous first-order differential equations,
\begin{equation}
\frac{d({u^2})}{d{\bar{\tau}}}
=\frac{2}{\gamma+1}c_s^2
\left[\frac{g_1^{\prime}}{g_1}
-\frac{1}{g_2}\left(\frac{\partial{g_2}}{\partial{r}}\right)\right]
\frac{f^{\prime}}{f}
\label{mod7}
\end{equation}
and
\begin{equation}
\frac{dr}{d{\bar{\tau}}}=
\frac{1}{1-u^2}\left(1-\frac{2}{\gamma+1}\frac{c_s^2}{u^2}\right)
+\frac{1}{\gamma+1}\frac{c_s^2}{g_2}\left(\frac{\partial{g_2}}{\partial{u^2}}\right)
\label{mod8}
\end{equation}
with ${\bar{\tau}}$  being an arbitrary parameter. In the two equations above 
$\delta c_s^2$ can be closed in terms of $\delta u^2$ and $\delta r$ 
with the help of (\ref{eq110a}). Having done so, one could then make use of 
solutions of the form, $\delta r \sim \exp ({\bar{\Omega}} \tau)$ and 
$\delta u^2 \sim \exp ({\bar{\Omega}} \tau)$,  
from which, ${\bar{\Omega}}$ would give the eigenvalues --- growth rates of 
$\delta u^2$ and $\delta r$ in ${\bar{\tau}}$ space --- of the stability matrix 
implied by (\ref{mod7}-\ref{mod8}). Detailed calculations will show the eigenvalues 
to be 
\begin{equation}
{\bar{\Omega}}^{2}=\left|{\bar{\beta}}^{4}c_{s}^{4}\chi_1^{2}+\xi_{1}\xi_{2}\right|_{\rm r=r_c}
\label{mod9}
\end{equation}
where ${\bar{\beta}}^2=\frac{2}{\gamma+1}$ and $\chi_1,\xi_1$ and $\xi_2$ can be 
expressed as polynomials of $r_c$.
${\bar{\Omega}}^2$ can be evaluated
for any \eker once the value of the corresponding critical point $r_c$ is known.
The structure of (\ref{mod9}) immediately shows that the only admissible 
critical points in the conserved Kerr system will be either saddle points 
or centre type points.
For a saddle point, ${\bar{\Omega}}^2 > 0$, while for a centre-type point,
${\bar{\Omega}}^2 < 0$.

For multi-critical flow characterized by a specific set of \eker, one can 
obtain the value of ${\bar{\Omega}}^2$ to be positive for the inner and the outer critical points,
showing that those critical points are of saddle type in nature. ${\bar{\Omega}}^2$
comes out to be negative for the middle critical point, confirming that the middle critical
point is of centre type and hence no transonic solution passes through it. One
can also confirm that all mono-transonic 
flow (flow with a single critical point characterized by $\left[{\cal E},\lambda\right]$
taken either from {\bf I} or from {\bf O} region) 
corresponds to saddle type critical point.
\subsection{Distinction between the two different category of flows, both having
three distinct critical points} 
\label{subsection4.4}
There are distinct topological differences between the multi-transonic flow characterized by 
$\left[{\cal E},\lambda\right]$ taken from the region marked by {\bf A}, and the 
region marked by ${\bf A_1}$. For region marked by {\bf A}, the 
entropy accretion rate ${\dot {\Xi}}$ for flows passing through the 
inner critical point is {\it greater} than that of the outer critical point
\begin{equation}
{\dot {\Xi}}\left(r_c^{in}\right)>{\dot {\Xi}}\left(r_c^{out}\right) 
\label{eqN2}
\end{equation}
while for the region marked by ${\bf A_1}$, the following relation holds
\begin{equation}
{\dot {\Xi}}\left(r_c^{in}\right)<{\dot {\Xi}}\left(r_c^{out}\right)
\label{eqN3}
\end{equation}
The above two relations show that  $\left[{\cal E},\lambda\right]$ region 
marked by {\bf A} represents the possibility of having 
multi-critical multi-transonic accretion,
while $\left[{\cal E},\lambda\right]{\in}\left[{\cal E},\lambda\right]_{\bf A_1}$
corresponds to the mono-transonic accretion, with an additional pair of 
critical points. 
For the multi-critical accretion, as we will see in more detail in the
subsequent sections, and as is illustrated in the panel diagrams (ii) - (iv) 
of figure 2, one can have a possibility that the
global accretion solution may pass through the outer critical/sonic
point, and a local accretion solution passing through the inner critical/sonic point, which 
folds back onto itself (at a point of inflexion formed at a certain radial distance 
$r{\equiv}r\left({\cal E},\lambda,\gamma,a\right)$), forming a 
{\it homoclinic orbit}\footnote{In the terminology of the autonomous dynamical systems,
a homoclinic orbit on a phase portrait is a solution that connects a
saddle type critical point to itself. See, e.g., \cite{js99,diff-eqn-book}
and the references therein
for further detail.} passing through the saddle type inner critical point
and embracing the centre type middle critical point flanked by the saddle type 
outer and the inner critical points. Such a homoclinic orbit can 
be connected with the standard non-homoclinic path through a standing 
shock, and hence, for a suitable initial boundary condition determined by 
$\left[{\cal E},\lambda,\gamma,a\right]$, accretion flow can pass through 
both the inner and the outer saddle type critical/sonic points. Hence  we
designate such flows (characterized by the set of parameters taken from the 
region {\bf A} of the parameter space, as shown in figure 1), which can have 
more than one sonic point available to it to pass through (in reality whether 
it will pass through both the sonic points depends on the possibility of the 
shock formation, see subsequent sections for further details) 
to be `multi-transonic' accretion.

The region {\bf A}, thus, as we will further discuss in the subsequent sections in even more
details, can further be divided into two regions. For the region where three 
critical points exist but no shock solution is allowed, the accretion is of the 
category multi-critical mono-transonic in strict sense, and such a region will hereafter be marked
as ${\bf A_{\sf NS}}$ -- which reads `multi-critical mono-transonic accretion 
flow with no shock'. For the rest of the {\bf A} region, a steady standing Rankine Hugoniot kind of 
shock forms, and accretion passes through both of the outer and the inner sonic points. 
Accretion is thus truly multi-transonic, and such a region in parameter space 
will be marked as ${\bf A_{\sf S}}$, which reads `multi-critical multi-transonic accretion 
with shock'.

On the other hand, flows described by the parameters taken from the region ${\bf A_1}$
of the parameter space behaves in a completely different way. 
Here, the global accretion solution passes through the inner critical/sonic 
point and hence it is mono-transonic. The homoclinic orbit is formed through 
the outer critical point, and the accretion sector of such homoclinic orbit 
can not be connected with the corresponding standard transonic accretion 
solution passing through the inner sonic point, by a shock (or through by 
whatever physical means), since the corresponding segment 
(lying in the length scale spanning the region between the outer critical point 
and the point of inflexion of the homoclinic loop) of the global transonic 
solution remains subsonic (shock can not form in a subsonic flow).
We thus call such flow topologies as the `mono-transonic accretion with an 
additional pair of critical points'. 

It is to be noted that the above mentioned solutions are commonly 
known as the 'multi-transonic wind' in the literature, owing to the fact that 
for a limited subset of values of $\left[{\cal E},\lambda,\gamma,a\right]$ 
taken from the region ${\bf A_1}$, the wind segment of the homoclinic orbit 
can be connected to the non homoclinic wind like solutions (passing through the 
inner sonic point) through a 
standing shock. However, we would not like to use such terminology here. 
Such terminology, as we feel, is misleading in the sense that the wind solution have 
no role in studying the accretion phenomena (except, perhaps, for studying the 
accretion powered outflow), and hence the term `multi-transonic wind' 
sounds as if there is no accretion taking place for the initial boundary conditions 
chose from the region ${\bf A_1}$, and such nomenclature is of compromised clarity only.

The actual situation, in contrary, is that there are several phases of 
transonic accretion flow, as we see in the subsequent section, and we concentrate 
only on the accretion part when we study accretion onto the black holes. 
The only true multi-transonic accretion are obtained if and only if shocks are formed.
Otherwise, the accretion is {\it always} mono-transonic, whether the flow possesses more 
than one critical points or not. Either such mono-transonic 
solution has only one saddle type critical point, and one corresponding sonic 
point (outer or inner), or it can have two sonic points available, and a 
homoclinic orbit passing through the inner critical point, but due 
to the absence of the shock formation phenomena, the transonic flow is
realizable only through the outer sonic point, and the flow is essentially 
topologically isomorphic to the mono-critical mono-transonic accretion from the 
region marked by {\bf O} in the parameter space as shown in the figure 1, or,
in its last variant, the flow can have three critical points, but the 
transonic accretion passes through only through the inner critical/sonic  point, 
accompanied by a homoclinic loop through the outer  critical point, and the 
corresponding accretion flow configuration is isomorphic to the 
mono-critical mono-transonic accretion passing through the inner critical/sonic point, 
as observed for flows chosen from the region marked by {\bf I} in the 
parameter space as shown in the figure 1. All such issues will be 
further clarified in greater detail in the next section.

The boundary between the region marked by {\bf A} and ${\bf A_1}$ is a very special one,
since for $\left[{\cal E},\lambda\right]$ defined by the boundary, the entropy 
accretion rate corresponding to the solution passing through the outer critical point
is exactly same as that of through the inner critical point
(${\dot {\Xi}}\left(r_c^{in}\right)={\dot {\Xi}}\left(r_c^{out}\right)$).
$\left[{\cal E},\lambda\right]$ defined by this boundary (rather $\left[{\cal E},\lambda,\gamma\right]$
for a fixed $a$, defined by this surface in general) provides the 
non-removable degenerate bistable/unstable solutions, we will discuss this issue in more 
details later.

There are other regions for $\left[{\cal E},\lambda\right]$ space for which either 
no critical points are formed, or two critical points are formed. These regions 
are not shown in the figure, since none of these regions is of our interest.
If no critical point is found, it is obvious that 
transonic accretion does not take place for those set of $\left[{\cal E},\lambda\right]$.
For two critical point region, one of the critical points are always of centre type, since 
according to the standard dynamical systems theory two successive critical points can 
not be of same type (both saddle, or both centre). Hence the solution which passes through the 
saddle type critical point would encompass the centre type critical point by forming a 
loop 
like homoclinic orbit,
and hence such solution would not be physically acceptable accretion solutions
since
such solutions do not connect infinity to the event horizon.
In subsequent sections, we will thoroughly describe the methodology 
for obtaining such flows parametrized by 
$\left[{\cal E},\lambda,\gamma,a\right]$ taken from the 
regions {\bf A} and ${\bf A_1}$.
\begin{figure}
\begin{center}
\includegraphics[scale=1.0, angle=0]{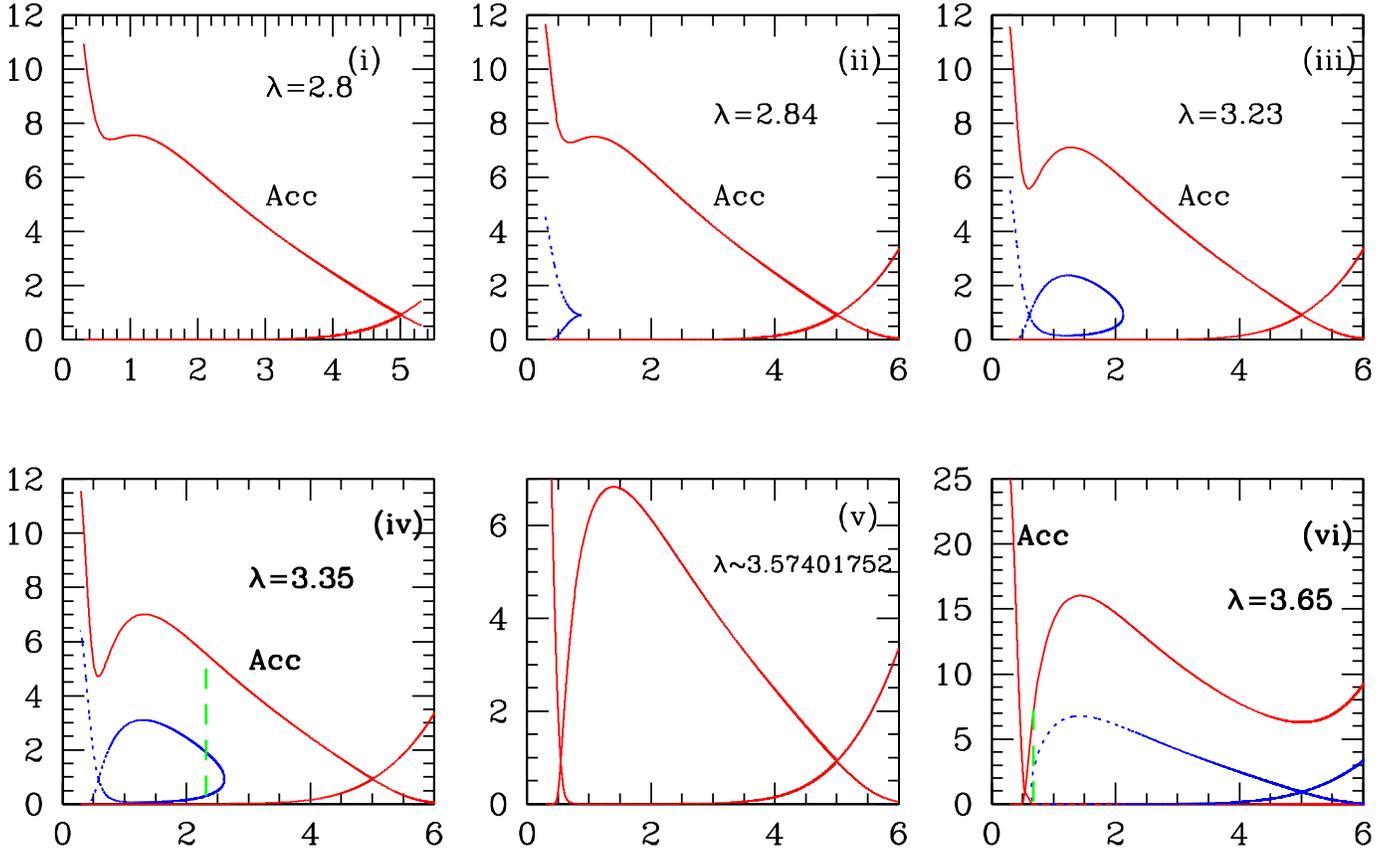}
\caption{\label{f2} \small{
Solution topologies for various cases as obtained by gradually increasing the 
flow angular momentum, by keeeping the other initial boundary conditions at a
fixed value of $\left[{\cal E}=1.0000033,\gamma=4/3,a=0.3\right]$. The usual accretion 
branch is denoted by solid line (red coloured in the online version) marked by 
`Acc', and the other solid line intersecting the accretion branch 
(at the outer critical points for panel (ii) - (iv) and at the inner critical 
point at panel (vi)) represents the corresponding wind solution. The homoclinic
orbit passing through the inner (for panel (ii) - (iv) ) or the outer (for panel (vi)) 
critical point is shown by the dotted line (blue coloured in the online version). 
The shock transition (at panel (iv) for true multi-transonic accretion, and at 
panel (vi) for shock formation in wind solution) is shown by vertical long dashed 
line (green coloured in the online version).
}}
\end{center}
\end{figure}

\section{Effect of change of the control parameter on the phase portrait}
\label{section5}
\subsection{Mono-transonic flow: Characteristics of the integral curves and 
the methodology to obtain the flow topology}
\label{subsection5.1}
\noindent
As has already been mentioned, we choose an $\left[{\cal E}-\lambda\right]$ 
projection of the entire four dimensional parameter space. The transonic flow properties and the 
corresponding flow topologies depends on four parameters of the \eker set. We would like to
illustrate such dependence by varying one parameter while keeping the 
other three parameters fixed. We will smoothly vary the angular momentum $\lambda$ of the flow and 
will observe its consequence on the phase topology. 
The corresponding topologies have been shown in six different panels (i) -- (vi) 
in figure 2. 
All the panel figures are drawn by plotting the Mach number along the
{\sf Y} axis and the distance (measured from the event horizon and scaled in the
unit of the gravitational radius) along the {\sf X} axis in logarithmic scale.
For all the topologies in the figure,
the value of ${\cal E}$ has been 
chosen to be $1.0000033$, which is in accordance with the value of the Bernoulli's constant 
for accretion flow onto the supermassive black hole located at our Galactic centre
\cite{monika}. The value of $\gamma$ has routinely been taken 
to be $4/3$, and $a=0.3$ has been taken as a representative value of the 
Kerr parameter. Same procedure can be performed using any other value 
of $a$ as well. Stable mono-transonic global solutions through the outer critical point 
is available for $0 < \lambda {\le} \lambda_1$, where $\lambda_1$ has the value in between
2.839 and 2.84. For such an interval of $\lambda$, the solution topology has been 
shown in the panel (i) of the figure 2, hereafter described as panel (ii). 

We now describe (in somewhat great detail) the methodology for obtaining a transonic 
solution topology on a phase portrait. This requires numerical integration 
of the equations describing the velocity gradient, as well as the gradient of the 
acoustic velocity, of the accretion flow. One thus needs to define the value of
both the velocity gradients at the critical point, pick up the required quantities (the 
value of the dynamical and the acoustic velocities, and their 
respective gradients at the critical 
point, and of the critical point itself), and to use those values 
as the starting values for the integration, like what is to be done in case of 
the standard initial value problem corresponding to the numerical solution of 
the differential equation. 

To obtain the dynamical velocity gradient at the 
critical point, one applies l'Hospital's rule on (\ref{eq110}). 
After some algebraic manipulations,
the following quadratic equation
is formed,
which can be solved
to obtain $(du/dr)_{\vc}$ (see \cite{bdw04} for further details):
\begin{equation}
\alpha \left(\frac{du}{dr}\right)_{\vc}^2 + \beta \left(\frac{du}{dr}\right)_{\vc} + \zeta = 0,
\label{eq113}
\end{equation}
\noindent
where the coefficients are:
\begin{eqnarray}
\alpha=\frac{\left(1+u^2\right)}{\left(1-u^2\right)^2} - \frac{2\delta_1\delta_5}{\gamma+1}, 
 \quad \quad \beta=\frac{2\delta_1\delta_6}{\gamma+1} + \tau_6,
 \quad \quad \zeta=-\tau_5;
& & \nonumber \\
\delta_1=\frac{c_s^2\left(1-\delta_2\right)}{u\left(1-u^2\right)}, \quad \quad
\delta_2 = \frac{u^2 v_t \sigma}{2\psi}, \quad \quad
\delta_3 = \frac{1}{v_t} + \frac{2\lambda^2}{\sigma} - \frac{\sigma}{\psi} ,
\quad \quad \delta_4 = \delta_2\left[\frac{2}{u}+\frac{u v_t \delta_3}{1-u^2}\right],
& & \nonumber \\
~
\delta_5 = \frac{3u^2-1}{u\left(1-u^2\right)} - \frac{\delta_4}{1-\delta_2} -
           \frac{u\left(\gamma-1-c_s^2\right)}{a_s^2\left(1-u^2\right)},
\quad \quad \delta_6 = \frac{\left(\gamma-1-c_s^2\right)\chi}{2c_s^2} +
           \frac{\delta_2\delta_3 \chi v_t}{2\left(1-\delta_2\right)},
& & \nonumber \\
\tau_1=\frac{r-1}{\Delta} + \frac{2}{r} - \frac{\sigma v_t\chi} {4\psi},
\quad \quad
\tau_2=\frac{\left(4\lambda^2v_t-a^2\right)\psi - v_t\sigma^2} {\sigma \psi},
& & \nonumber \\
\tau_3=\frac{\sigma \tau_2 \chi} {4\psi},
\quad \quad
\tau_4 = \frac{1}{\Delta} 
       - \frac{2\left(r-1\right)^2}{\Delta^2}
       -\frac{2}{r^2} - \frac{v_t\sigma}{4\psi}\frac{d\chi}{dr},
& & \nonumber \\
\tau_5=\frac{2}{\gamma+1}\left[c_s^2\tau_4 -
     \left\{\left(\gamma-1-c_s^2\right)\tau_1+v_tc_s^2\tau_3\right\}\frac{\chi}{2}\right]
   - \frac{1}{2}\frac{d\chi}{dr},
& & \nonumber \\
\tau_6=\frac{2 v_t u}{\left(\gamma+1\right)\left(1-u^2\right)}
       \left[\frac{\tau_1}{v_t}\left(\gamma-1-c_s^2\right) + c_s^2\tau_3\right].
\label{eq114}
\end{eqnarray}
Note that all the above quantities are evaluated at the critical point.

Hence we compute the critical advective velocity gradient as
\begin{equation}
\left(\frac{du}{dr}\right)_{\rm r=r_c}
=-\frac{\beta}{2\alpha}
{\pm}
\sqrt{\beta^2-4\alpha{\zeta}}
\label{eq113a}
\end{equation}
where the `+' sign corresponds to the accretion solution and
the `-' sign corresponds to the wind solution, see the
following discussion for further details.
Similarly, the space gradient of the acoustic velocity
$dc_s/dr$ and its value at the critical point has also been
calculated.

For the solution topologies shown in the figure 2,
using the specified set of \eker, we first solve
the equation for the energy first integral defined at the 
critical point, to obtain the
corresponding critical point $r_c= 99851.8623$, which is the point of 
intersection in between the accretion branch marked by `Acc' in the figure, and
the corresponding wind branch, the other curve shown in the figure. 
We then calculate the critical value of the advective
velocity gradient at $r_c$ from (\ref{eq113a}). 
By integrating  the equation for the dynamical and the acoustic flow 
velocity gradient (Eq. (\ref{eq110a}) and (\ref{eq110}), respectively),
from the critical point, using the fourth-order Runge-Kutta method,
we then calculate the
local advective velocity, the polytropic sound speed,
the Mach number, the fluid density, the disc height, the bulk temperature of the
flow, and any other relevant dynamical and thermodynamic quantity
characterizing the flow.
In this way we
obtain the accretion branch `Acc' by employing the above mentioned
procedure.

The solution, as is obvious from the figure, is two-fold
degenerate owing to the $\pm u$ degeneracy which
reflects the physical accretion/wind degeneracy.
We have, however, removed the degeneracy by orienting the curves, and thus
each line represents either the wind or accretion.
 We have arbitrarily assigned  the $+$ sign solution in Eq. (\ref{eq113a}) to the
  accretion
and the $-$ sign solution in Eq. (\ref{eq113a}) to
the `wind' branch. This wind branch is just a
mathematical counterpart of the accretion solution (velocity reversal
symmetry of accretion),
owing to the presence of the quadratic term
of the dynamical velocity in the equation governing the
energy momentum conservation.

The term `wind solution' has
a historical origin.
The solar wind solution first introduced
by Parker  ~\citep{parker}
has the same  topology profile as that of the
wind solution obtained in classical Bondi accretion ~\citep{bondi}. Hence the
name `wind solution' has been adopted in a more general sense.
The wind solution thus represents a hypothetical process,
in which, instead of starting from infinity
and heading towards the black hole, the flow
 generated near the black-hole event horizon would fly away from the
black hole towards infinity. The topology of such a process
is represented by the wind solution in the panel, which, as already been mentioned,
is the curve intersecting the accretion branch `Acc' at the outer critical point.

The above procedure for obtaining the flow topology
is also applied to
draw the mono-transonic accretion/wind branch
through the inner sonic point.
In fact, the same procedure may be
used to draw real physical transonic accretion/wind solutions passing through any
acceptable saddle-type critical point.
Note, however,
that the sector of the accretion branch starting from the 
intersection of the accretion and the wind solution in the panel diagram
and ending at the event horizon
is {\it not} the complete
subsonic branch, because the point of intersection is a critical point and not a sonic point.
Using the procedure described above
we have to integrate the flow from the critical point to the
sonic point
where the Mach number becomes unity.
\subsection{Appearance of the homoclinic orbit and the emergence of the 
multi-transonicity}
\label {subsection5.2}
\noindent
We now further increase $\lambda$ so as to reach a value $\lambda_2$ -- greater than 
but infinitesimally close to $\lambda_1$, i.e.:
$$
{\large\sf Lt}_{{\epsilon}{\rightarrow}0}~|\lambda_2-\lambda_1|=\epsilon
$$
$\lambda_2$ thus characterizes the flow topology generated at the 
right side of the boundary between the {\bf O} and the {\bf A} regions 
shown in figure 1, and infinitesimally 
close to it. Such solution has been shown in the panel (ii). 
The line separating the {\bf O} and the {\bf A} region is actually the 
line of bifurcation of the critical point, because 
the number of critical point increases after crossing such line.
The line of separation between the {\bf O - A} region, in general, 
is a three dimensional hypersurface for the 
generic multi dimensional parameter space spanned by \eker. Additional 
critical points appear in pairs once such a line is crossed from the left side
(by gradually increasing the value of the flow 
angular momentum $\lambda$) of the parameter space. Hence, while increasing 
$\lambda$ smoothly, initially there is a region of a single saddle point,
then followed by the pair creation of a saddle and a centre type point after 
crossing the {\bf O - A} boundary. That is what is manifested in panel (ii). In addition to 
a transonic accretion solution passing through the outer critical point, which is denoted by 
the solid lines (red coloured in the online version)
a tiny loop of homoclinic orbit (shown using dotted line, blue coloured in the 
online version) appears representing the flow through 
(and around) a nascent saddle-centre pair of critical points. Physically, this is what it 
has to be, because with a centre type point but without another saddle 
point, the flow solutions will all curl about the centre type point, and there
will be no means of connecting the event horizon with infinity through an 
isolated solution passing through the saddle type inner critical point, 
and embracing the centre type middle critical point 
(here, the inner and the middle critical points are formed at 7.2 and 7.4 $r_g$, respectively).
The homoclinic path has its existence in isolation, and unlike the
transonic solution passing through the outer critical point, can not be 
regarded as a global transonic solution on its own. A physically 
acceptable flow through the inner critical point has to be 
connected with the solution passing through the outer critical point.
As we will see in the subsequent paragraphs, a standing Rankine-Hugoniot
shock perfectly accomplish the task. 

Accretion flow through the inner critical point has higher value of 
the entropy accretion rate compared to that of passing through the 
outer critical point. The ratio ${\sf R}_{\dot \Xi}$ of 
${\dot \Xi}_{in}$ and ${\dot \Xi}_{out}$ attains its maximum
value close to the {\bf O -- A} boundary, and then monotonically decreases as one smoothly transits 
to the higher end of $\lambda$ (toward the boundary separating the 
{\bf A} and the ${\bf A_1}$ regions, respectively).

As we further process toward higher value of $\lambda$, the area enclosed within the
homoclinic orbit starts increasing and finally it fills up the entire 
region bounded by the transonic accretion and wind passing through the 
outer critical point. The situations have been 
demonstrated sequentially in the successive panel figures (iii), (iv) and (v). 
\subsection{Multi-critical  accretion with and without shock}
\label{subsection5.3}
\noindent
In panel (iii), ${\sf R}_{\dot \Xi}$ is 212.52. Shock, however, does not 
form even if three critical points are present. Such multi-critical flow 
(which is actually a mono-transonic accretion because in absence of the 
shock, accretion flow can pass only through the outer sonic point even if the inner 
sonic point formally exists) without a standing shock is observed for a range of value of 
$\lambda_2{\le}\lambda{\le}\lambda_3$ where $\lambda_3{\sim}3.23{\dot 9}$, 
which defines the boundary of the ${\bf A_{\sf NS}}$ region.

As $\lambda$ is further increased so that $\lambda>\lambda_3$, Rankine-Hugoniot 
conditions (relativistic Rankin-Hugoniot conditions are explicitly derived in 
the next section) gets satisfied, and standing shock forms. Panel (iv) shows the 
topology of the multi-transonic accretion with shock 
solutions, where the shock transition has been 
marked by the vertical long-dashed (green coloured in the online 
version) line segment. The shock location, shock strength, in which is defined as the 
pre (-) to the post (+) shock ratio (defined not in the temporal 
sequence but in the spatial sequence, of the Mach number) and is denoted by 
$M_-/M_+$, and the entropy enhancement 
ratio at the shock ${\sf R}_{\dot \Xi}$ are found to be
18.3336735 and 74.9621923, respectively.

It is to be noted that the panel (iv) is not the solution corresponding to the 
minimum value of $\lambda$ for which the shock forms, i.e., does not 
correspond to the value of $\lambda$ separating the ${\bf A_{\sf NS}}$ and 
the ${\bf A_{\sf S}}$ region.
The strongest shock (forms for the maximally allowed value of ${\sf R_{\dot \Xi}}$
in ${\bf A_{\sf S}}$) are formed for $\lambda_4$ where
$$
{\sf Lt}_{\epsilon{\rightarrow}0}|\lambda_4-\lambda_3|=\epsilon
$$ 
For example, for $\lambda=3.24$, the shock forms at 27.5$r_g$, and the 
corresponding shock strength and the entropy enhancement ratio at the
shock are found to be 41.7 and 197.62, respectively. 

The relativistic shock equations and its derivation, along with a detail 
description of the methodology of finding the multi-transonic flow 
with and without shock is presented in the subsequent paragraphs. However,
from panel (iii) and (iv), it is to be understood that shock does not necessarily 
form for all multi-critical  flow. Having two saddle type critical 
points is thus a necessary but not a sufficient condition. The 
reason behind this disparity will be analytically justified in the 
section (\ref{section6}) and section (\ref{section7})
\subsection{Methodology for obtaining multi-transonic topology}
\label{subsection5.3}
Using a specific set of \eker,
one first solves the equation for ${\cal E}$ at the critical point 
to find out the corresponding three
critical points, saddle type inner, centre type middle,  and the
saddle type outer. The space gradient of the flow velocity as well as the
acoustic velocity at the saddle type critical point is then obtained.
Such $u_{\vc},{c_s}_{\vc}$,$dc_s/dr_{\vc}$  and
$du/dr_{\vc}$ serve as the initial value condition for performing the numerical
integration of the advective velocity gradient using the fourth-order
Runge-Kutta method. Such integration provides the outer {\it sonic} point
(located closer to the black hole compared to the outer critical 
point, since the Mach number at the outer critical point is less 
than unity), the
local advective velocity, the polytropic sound speed,
the Mach number, the fluid density, the disc height, the bulk temperature of the
flow, and any other relevant dynamical and thermodynamic quantity
characterizing the flow. The corresponding wind solution through the outer sonic
point is obtained in the same way, by taking the other value (note that being a quadratic
algebraic equation, solution of (\ref{eq113a}) provides two initial values, one for the 
accretion and the other for the wind) of $du/dr_{\vc}$.

The respective
accretion and the wind solutions passing through the inner critical point 
are obtained following exactly the same procedure as has been
used to draw the accretion and wind topologies passing through the
outer critical point. Note, however, that the accretion solution through the inner critical point
folds back onto the wind solution which is a homoclinic orbit through the 
inner critical point
encompassing  the centre type middle critical point.
A physically acceptable  
transonic solution must be globally consistent, i.e. it must connect
the radial infinity
with the black-hole event horizon.
Hence, for multi-transonic accretion, there is no individual existence
of physically acceptable accretion/wind solution passing through the inner
critical (sonic) point, although, depending on the initial boundary conditions,
such solution may be clubbed with the
accretion solution passing through outer critical point, through a standing shock.

The set $\left[{\cal E},\lambda\right]_{\bf A}$
(or more generally $\left[{\cal E},\lambda,\gamma,a\right]_{\bf A}$)
thus
produces doubly degenerate accretion/wind solutions.
Such two
fold degeneracy may be removed by the entropy considerations since
the entropy accretion rates for solutions passing through the inner critical point
and the outer critical point are generally not equal.
For any $\left[{\cal E},\lambda,\gamma,a\right]{\in}
\left[{\cal E},\lambda,\gamma,a\right]_{\bf A}$
we find that the entropy accretion rate  ${\dot \Xi}$ evaluated for the
complete accretion solution passing through the outer critical point
is less than that of the rate  evaluated for the incomplete accretion/wind solution
passing through the inner critical point.
Since the quantity ${\dot \Xi}$
is a measure of the specific entropy density of the flow,
the solution passing through the outer critical point will naturally tend
to make
a transition to its higher entropy counterpart,
i.e. the globally incomplete accretion solution
passing through the inner critical point.
Hence, if there existed a mechanism for
the accretion solution passing through the outer critical point
to increase
its entropy accretion rate exactly by an amount
\begin{equation}
{\Delta}{\dot \Xi}=
{\dot \Xi}(r_c^{\rm in})-{\dot \Xi}(r_c^{\rm out}),
\label{eq48}
\end{equation}
there would be a transition to the
accretion solution
passing through the
inner critical point.
Such a transition would take place at
a radial distance somewhere  between the radius of the inner  sonic point
and the
radius
of the point of inflexion of the homoclinic orbit.
In this way one would obtain a true multi-transonic accretion solution connecting
the infinity and the event horizon,
which includes a part of the accretion
solution passing through the inner critical, and hence the inner sonic point.
One finds that for some specific values of
$\left[{\cal E},\lambda,\gamma,a\right]_{\bf A}$,
a standing Rankine-Hugoniot shock may accomplish this task.
A supersonic accretion through the outer {\it sonic} point 
(which in obtained by integrating the flow starting from the outer
critical point)
can generate
entropy through such a shock formation and can join the flow passing through
the inner {\it sonic} point 
(which in obtained by integrating the flow starting from the outer
critical point). Below we will carry on a detail discussion on such
shock formation.
\subsection{The shock formation and related phenomena}
\label{subsection5.4}
In the present work, the basic equations governing the flow
are the energy and baryon number
conservation equations which contain no dissipative
terms and the flow is assumed to be inviscid.
Hence, the shock
which may be produced in this way can only be of Rankine-Hugoniot type
which conserves energy. The shock thickness must be relatively small
in this case, otherwise non-dissipative
flows may radiate energy through the upper and the lower boundaries because
of the presence of strong temperature gradient in between the inner and
outer boundaries of the shock thickness.
In the  presence of a shock
the flow may have the following profile.
A subsonic flow starting from infinity first becomes supersonic after crossing
the outer sonic point and somewhere in between the outer sonic point and the inner
sonic point
the shock transition takes place and forces the solution
to jump onto the corresponding subsonic branch. The hot and dense post-shock
subsonic flow produced in this way becomes supersonic again after crossing
the inner sonic point and ultimately dives supersonically into the
black hole.
A flow heading towards a neutron star can have the liberty of undergoing
another shock transition
after it crosses the inner sonic point.
Or, alternatively, a shocked flow heading towards a neutron star
need not encounter the inner sonic point at all. This is because,
the hard surface boundary
condition of a neutron star by no means prevents the flow
from hitting the star surface subsonically.

For the complete general relativistic accretion flow discussed in this article,
the energy momentum tensor ${\large\sf T}^{{\mu}{\nu}}$, the four-velocity $v_\mu$,
and the speed of sound $c_s$ may have discontinuities at a
hypersurface $\Sigma$ with its normal $\eta_\mu$.
 Using the energy momentum conservation and the
continuity equation, we have:
\begin{equation}
\left[\left[{\rho}v^{\mu}\right]\right] {\eta}_{\mu}=0,
\left[\left[{\large\sf T}^{\mu\nu}\right]\right]{\eta}_{\nu}=0.
\label{eqa51}
\end{equation}
For a perfect fluid, we thus formulate the relativistic
 Rankine-Hugoniot conditions as
\begin{equation}
\left[\left[{\rho}u\Gamma_{u}\right]\right]=0,
\label{eqa52}
\end{equation}
\begin{equation}
\left[\left[{\large\sf T}_{t\mu}{\eta}^{\mu}\right]\right]=
\left[\left[(p+\epsilon)v_t u\Gamma_{u} \right]\right]=0,
\label{eqa53}
\end{equation}
\begin{equation}
\left[\left[{\large\sf T}_{\mu\nu}{\eta}^{\mu}{\eta}^{\nu}\right]\right]=
\left[\left[(p+\epsilon)u^2\Gamma_{u}^2+p \right]\right]=0,
\label{eqa54}
\end{equation}
where $\Gamma_u=1/\sqrt{1-u^2}$ is the Lorentz factor.
The first two conditions (\ref{eqa52})
and (\ref{eqa53})
are trivially satisfied owing to the constancy of the
specific energy and mass accretion rate.
The constancy of mass accretion yields
\begin{equation}
\left[\left[
K^{-\frac{1}{\gamma-1}}
\left(\frac{\gamma-1}{\gamma}\right)^{\frac{1}{\gamma-1}}
\left(\frac{c_s^2}{\gamma-1-c_s^2}\right)^{\frac{1}{\gamma-1}}
\frac{u}{\sqrt{1-u^2}}
H(r)\right]\right]=0.
\label{eqa55}
\end{equation}
The third Rankine-Hugoniot condition
(\ref{eqa54})
may now be  written as
\begin{equation}
\left[\left[
K^{-\frac{1}{\gamma-1}}
\left(\frac{\gamma-1}{\gamma}\right)^{\frac{\gamma}{\gamma-1}}
\left(\frac{c_s^2}{\gamma-1-c_s^2}\right)^{\frac{\gamma}{\gamma-1}}
\left\{\frac{u^2\left(\gamma-c_s^2\right)+c_s^2}{c_s^2\left(1-u^2\right)}\right\}
\right]\right]=0.
\label{eqa56}
\end{equation}
Simultaneous solution of Eqs. (\ref{eqa55}) and (\ref{eqa56}) yields the `shock invariant'
quantity
${\cal S_h}$ 
which changes continuously across the shock surface.
We obtain the analytical expression for ${\cal S_h}$ in terms of various local 
accretion variables and 
initial boundary conditions. 
The shock location in
multi-transonic accretion
is found in the following way.
While performing the numerical integration 
along the solution passing through the outer 
critical point, we calculate the shock invariant
${\cal S}_h$ in addition to
$u$, $c_s$ and $M$. We also calculate  ${\cal S}_h$
while integrating along the solution passing through the inner critical 
point, starting from the inner {\it sonic}
point up to the point of inflexion of the homoclinic orbit.
We then determine
the radial distance $r_{sh}$, where the numerical values of ${\cal S}_h$,
obtained by integrating the two different sectors described above, are
equal. Generally,
for
any value of \eker 
allowing shock
formation,  one finds two formal shock locations,
one located in 
between the outer and the middle
sonic point and the other one located 
in between the
inner and the middle sonic point.
The shock strength is different for the inner and 
for the outer shock. 
According to the standard
local stability analysis (\cite{yk95}),
for a multi-transonic accretion, one can show that
only the shock formed  between
the middle
and the outer sonic point is stable.
Hereafter, whenever we mention the shock
location, we  always refer
to the stable shock location only.

We find that the shock location  correlates with
$\lambda$.
This is obvious because the higher the flow
angular momentum, the greater the rotational energy content
of the flow. As a consequence, the strength of the centrifugal
barrier which is responsible to break the incoming flow by forming a shock
will be higher and
the location of such a barrier will be farther away from the
event horizon.
However, the shock location
anti-correlates with ${\cal E}$ and $\gamma$.
This  means that for the same ${\cal E}$ and $\lambda$, in the purely
non-relativistic flow the shock
will form closer to the black hole compared with
the ultra-relativistic flow. Besides, we find that the shock strength
anti-correlates with the shock location $r_{sh}$,
which indicates that
the
closer to the black hole the shock forms , the higher the strength 
and the entropy enhancement ratio are.
The ultra-relativistic flows
are supposed to
produce the strongest shocks.
The reason behind this is also easy to understand. The closer to the black hole the shock
forms, the higher the available gravitational
potential energy must be released, and the radial
advective velocity required to have a more vigorous shock jump will be larger.
Besides we note  that as  the flow gradually approaches its purely
non-relativistic limit,
the shock may form for lower and lower angular momentum,
which indicates that for purely non-relativistic
accretion, the shock formation may take place even for a quasi-spherical flow.
However, it is important to mention that
a shock formation will be allowed
not for every
$\left[{\cal E},\lambda,\gamma,a\right]{\in}
\left[{\cal E},\lambda,\gamma,a\right]_{\bf A}$,
The numerical value for ${\cal S_h}$ will be the same 
along both the accretion through the outer and the inner sonic point, only
at certain points (the shock locations)
for a specific subset of $\left[{\cal E},\lambda,\gamma,a\right]_{\bf A}$,
for which a steady, standing shock solution
will be found.

We further find that the shock location correlates with the black hole spin parameter, whereas the 
shock strength and the entropy enhancement ratio at the shock anti correlates 
with the Kerr parameter. The post to pre shock velocity correlates with the black hole 
spin, whereas the post to pre shock temperature, density and pressure anti correlates
with the Kerr parameter.

\subsection{The `asymmetrically hunched fish' and the heteroclinic orbits}
\label{subsection5.5}
As we further increase $\lambda$, the shock location recedes outward, and the 
shock strength as well as ${\sf R}_{\dot \Xi}$ decreases monotonically. 
Meanwhile, the homoclinic path embracing the middle critical point 
starts filling up the entire region accessible to it. Finally, when  
$\lambda$ approaches the value $\lambda_5{\sim}3.57401752$, the interface 
between the {\sf A} and the ${\sf A_1}$ regions is approached, and one obtains
the characteristic phase portrait as that of panel (v).

The above mentioned line of interface, as has already been discussed, represents the 
state where the entropy accretion rate for solution passing through the inner critical 
point is exactly the same for the solution passing through the outer critical point. 
Integral curves on the phase portrait are allowed to intersect only at the 
critical points. Hence two saddle points in this case will be connected through 
{\it heteroclinic orbits}. A heteroclinic orbit is generally a trajectory on a two 
dimensional phase portrait which connects one saddle type equilibrium points to the
other saddle type equilibrium point. If the two saddle points are the same, then it 
becomes a homoclinic orbit. Existence of the heteroclinic orbit is associated with
the stability issues in the dynamical systems of a two dimensional vector field
(velocity filed, for example). A heteroclinic orbit is contained in the stable 
manifold of one saddle point and in the unstable manifold of the other saddle 
point. A fundamental fact arising from the theory of the structural stability 
is that if two saddle type critical points are connected through the heteroclinic
orbit, then the local phase portrait near this orbit can be changed by an arbitrarily
small smooth perturbation. In effect, a perturbation can be chosen such that, in 
the phase portrait of the perturbed vector field, the saddle connection is 
broken. Thus, in particular, a vector field with two saddle 
points connected by a heteroclinic orbit is not structurally stable 
with respect to the class of all smooth vector fields (\cite{diff-eqn-book}).

Drawing the phase portrait for the above mentioned topology 
will entail the tuning of the flow parameters (and the initial boundary 
conditions) with infinite numerical precession. The solution topology 
looks like a fish, which has an asymmetrically placed hunch on its back.
The collection of all the $\left[{\cal E}-\lambda\right]$ pair, 
for which such `fish topologies' are produced, is thus a line 
characterizing the condition of the heteroclinicity.

\subsection{Topology from region ${\bf A_1}$ and the pair annihilation of the critical points}
\label{subsection5.3}
\noindent
A further increase of $\lambda$ allows to enter in the ${\sf A_1}$ region.
A representative diagram for such flow topology is shown in the panel (vi).
The converse behaviour (compared to the multi-transonic accretion) of the 
phase portrait for multi-transonic accretion) is quite obvious. here the governing principle,
as already been mentioned in Sect. \ref{subsection4.4}, is that the entropy accretion rate  pertaining to 
the inner critical point is greater than that of the outer critical point, and hence 
the homoclinic orbit passes through the outer critical point.

The topological flow profile for the present case is 
different from that of the multi-transonic accretion. Here the homoclinic orbit 
passes through the outer critical point. The procedure for finding the 
complete flow topology with shock is essentially same as 
that for finding the complete topology of a true multi-transonic shocked 
accretion flow as mentioned earlier.
To find the shock location,
here one calculate the numerical values of ${\cal S}_h$ along the
wind  solution passing through the inner critical point
are compare with the numerical values of ${\cal S}_h$
along the wind solution passing through the outer
critical point, and the stable shock location
is found accordingly. Extremely strong shocks are found to 
form very close to the black hole 
(on the off equatorial plane) for this situation. We will show in our future work that such
wind solutions with strong shocks are useful in explaining the 
GRB light curves (Janiuk, Czerny \& Das, in preparation). 

For higher value of $\lambda$, the homoclinic orbit through the outer critical point 
start shrinking, and finally disappear with pair-annihilation of the saddle type 
outer critical point and the centre type inner critical point once we cross the 
hypersurface separating the ${\sf A_1}$ and the mono-transonic accretion 
(passing only through the inner critical point) zone. All that is left 
behind is a single saddle type critical point, through which mono-transonic 
accretion and wind passes. 
\begin{figure}
\vskip -5.0truecm
\begin{center}
\includegraphics[scale=0.7, angle=0]{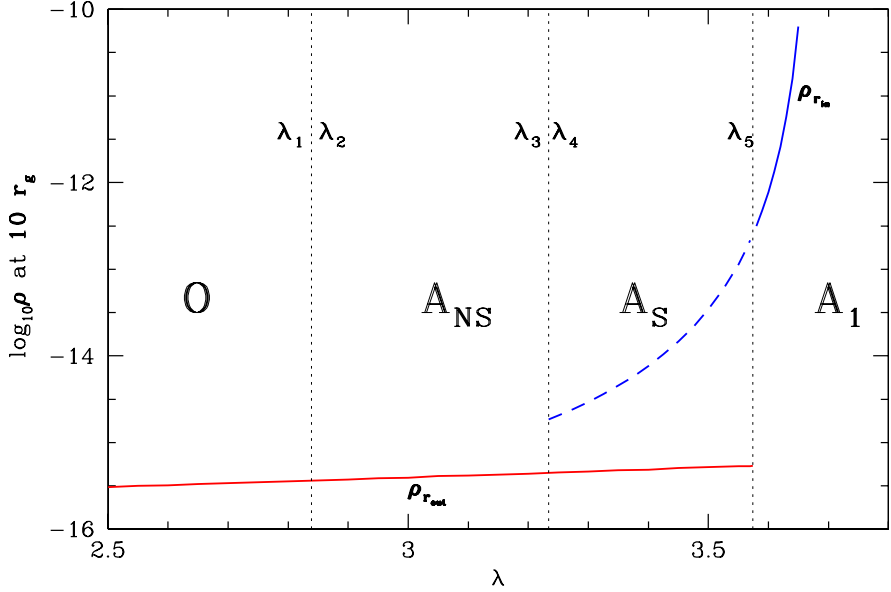}
\caption{\label{f3} \small{
In the figure, $\rho_{r_{out}}$ and $\rho_{r_{in}}$ represent
the density of the accreting material measured at 10$r_g$ along the 
transonic accretion passing through the outer and the inner sonic points,
respectively. The long dashed line corresponds to the range of the angular momentum
for which a steady standing Rankine Hugoniot shock forms. The figure has been drawn 
for $\left[{\cal E}=1.0000033,\gamma=4/3,a=0.3\right]$.
}}
\end{center}
\end{figure}

\section{Continuity in the parameter space and the hysteresis effect}
\label{section6}
\noindent
The heteroclinic fish like solution as described in the section 5.6 marks the transition between two different 
flow topologies, a multi-transonic accretion with a built-in homoclinic orbit (as shown in the successive 
panel figures (ii) - (iv)) and a mono-transonic accretion with an additional non accessible homoclinic
orbit (as shown in the panel figure (iv)). It is to be understood that the solutions with shock for the value of 
the flow angular momentum higher than $\lambda_5$ are of completely different character compared to the 
solution parametrized by the value of angular momentum lower than $\lambda_5$. We illustrate this issue through 
the following procedure:

We set $r_{10}=10r_g$ as our reference point.  For a transonic solution, we then calculate the density 
of the accerting matter at $r_{10}$ as a function of $\lambda$, keeping the other initial boundary 
conditions $\left[{\cal E},\gamma,a\right]$ fixed. This has to be done in several steps. For mono-transonic
accretion passing through the outer sonic point only, we can have only one choice to calculate 
$\rho$ at $r_{10}$, that is to calculate the density at $r_{10}$ only on the solution passing through the 
outer sonic point, $r_{out}$, and denoting it by $\rho_{r_{out}}$. We do it without any ambiguity
for a span of value of $\lambda$, starting from a very small value (almost equal to zero, just when the 
departure from the pure Bondi type accretion stars) upto the value $\lambda=\lambda_1$. Beyond $\lambda_1$,
for $\lambda>\lambda_1$, in principle there are two choices for the density calculation at $r_{10}$. One is along 
the flow passing through $r_{out}$, to obtain $\rho_{r_{out}}$ for the multi-transonic flow. Another option is
to compute the density $\rho_{r_{in}}$ along the multi-transonic accretion flow passing through the inner 
sonic point $r_{in}$. Both $\rho_{r_{out}}$ and $\rho_{r_{in}}$ can be calculated as a function of $\lambda$
for the span $\lambda_1<\lambda<\lambda_5$. However, since for the span $\lambda_1<\lambda<\lambda_3$ shock 
does not form, $\rho_{r_{in}}$ is a meaningless quantity to define for this span of $\lambda$, and thus
for any value of the angular momentum upto $\lambda=\lambda_3$, density at $r_{10}$ means $\rho_{r_{out}}$ only.
However, the situation is different for $\lambda_3<\lambda{\le}\lambda_5$, where both $\rho_{r_{out}}$ and $\rho_{r_{in}}$
can have feasible meaning. However, $\rho_{r_{out}}$ is practically meaningful if shock does not form
(since it has been thoroughly verified that the lowest possible shock location $r^{\sf minimum}_{\sf sh}$ for 
the entire range of our choice of $\left[{\cal E},\lambda,\gamma,a\right]$ is always greater than $r_{10}$), and
$\rho_{r_{in}}$ is practically meaningful only if the shock forms. 
Beyond $\lambda_5$, for $\lambda>\lambda_5$, once again one has mono-transonic accretion (with an additional 
homoclinic orbit of course), now passing through the inner sonic point $r_{in}$. Hence the measured density at $r_{10}$
for $\lambda>\lambda_5$ implies $\rho_{r_{in}}$ only. 

Hence, we have $\rho_{r_{out}}$ defined over the range for any value of the flow angular momentum 
upto $\lambda=\lambda_5$, and, on the contrary, $\rho_{r_{in}}$ is defined only over the range
$\lambda{\ge}\lambda_3$. In figure 3, we plot, in logarithmic scale,  
$\rho_{r_{out}}$ using the solid line (red coloured in the online 
edition) and $\rho_{r_{in}}$ using the solid line (in the region {\sf IV}) and the dashed line (in the region 
{\sf III}) respectively, as a function of the specific flow angular momentum $\lambda$.

The local density changes by a factor over 500 at $\lambda_5$ which well illustrates the global quantitative 
change in the flow. However, when we add the solutions passing through the inner and the outer sonic points with a shock
to the same plot, those solutions form a continuation of the
previous branch below  $\lambda_5$. The shock for $\lambda$ just below $\lambda_5$ is extremely weak
(with the corresponding shock strength having value very close to the unity),
and located infinitesimally close inward of the outer critical point. With a further decrease of $\lambda$
the shock strength measured as a ratio of the Mach numbers above and below the shock increases monotonically,
and the shock position moves inward.

The branch with a shock also does not cover the whole range of angular momentum values -- there is a minimum
value of $\lambda$ (for a given Bernoulli constant, the polytropic index and the black hole spin) 
for which the Rankine-Hugoniot
conditions (see Sect.~\ref{subsection5.4}) can be satisfied. This value, marked as $\lambda_4$ in 
the section \ref{subsection5.3}, describes
the discontinuous change in the solution along the branch with a shock with a further drop in the angular
momentum, as for still lower values of $\lambda$ only the solutions without shock exist. The local density
change in the flow at $r_{10} $ accompanying the slight change of angular momentum along this branch is by
a factor 4, much less than the factor 500 at the transition at $\lambda_5$ for the shock-free solutions.

Since the system is likely to choose a solution close to the previous one under the slight change of the
outer boundary conditions (here, the angular momentum), the choice between the solution with a shock and
without a shock for $\lambda_4 < \lambda <\lambda_5$ will depend on the previous history of the system.
If the initial value of the angular momentum is lower than $\lambda_4$ and slowly increases the system
will follow the lower branch ($\rho_{r_{out}}$) in Fig. 3 without a shock and will show a dramatic change when
passing from $\lambda < \lambda_5$ to  $\lambda > \lambda_5$. If, on the other hand, the system evolution
starts with  $\lambda > \lambda_5$ and the angular momentum of the infalling material slowly decreases
the system is more likely to follow the upper branch in Fig. 3, without any discontinuity in
the appearance down to $\lambda_4$ but the solution will display an increasingly stronger shock with a
quasi-stationary increase of the angular momentum of the matter.

We thus expect a hysteresis effect in the system behavior under the conditions of slowly varying initial
angular momentum of the accreting matter. This may in turn lead to some asymmetry in the lightcurves of
the systems accreting in such a mode. The local density at $10 r_g$ is an indicator of the system
luminosity (the synchrotron emission increases with the plasma density and magnetic field).
The increase in the local density indicates an increase in a source brightness. Since the
discontinuity in the density profile is larger when the angular momentum and consequently the density
increases (at $\lambda_5$) the brightening should be faster/stronger than the subsequent faintenning
of the source (evolution along the upper branch ($\rho_{r_{in}}$) in Fig. 3, down to $\lambda_4$). However,
detailed study of this effect and comparison with observational data is beyond the scope of
the present paper.

In this context, it is to be noted here that the shock strength, which correlates with the 
entropy enhancement ratio ${R_{\dot {\Xi}}}$ at the shock, monotonically increases with decreasing 
$\lambda$, and at the same time anti-correlates with the shock location itself. The variation of the 
shock strength (solid line marked by $M_-/M_+$) and ${R_{\dot {\Xi}}}$ (dotted line marked by 
${R_{\dot {\Xi}}}$) with $\lambda$ is shown in the figure 4. The strongest shock thus forms closer to the 
black hole. From observational data, it is expected that the QPO profile of the SgrA* black hole corresponding to a 
small length scale shock formation. Shocks formed for the angular momentum $\lambda{\sim}\lambda_4$ thus, 
might corresponds to such QPO behaviour. 
\begin{figure}
\vskip -5.0truecm
\begin{center}
\includegraphics[scale=0.7, angle=0]{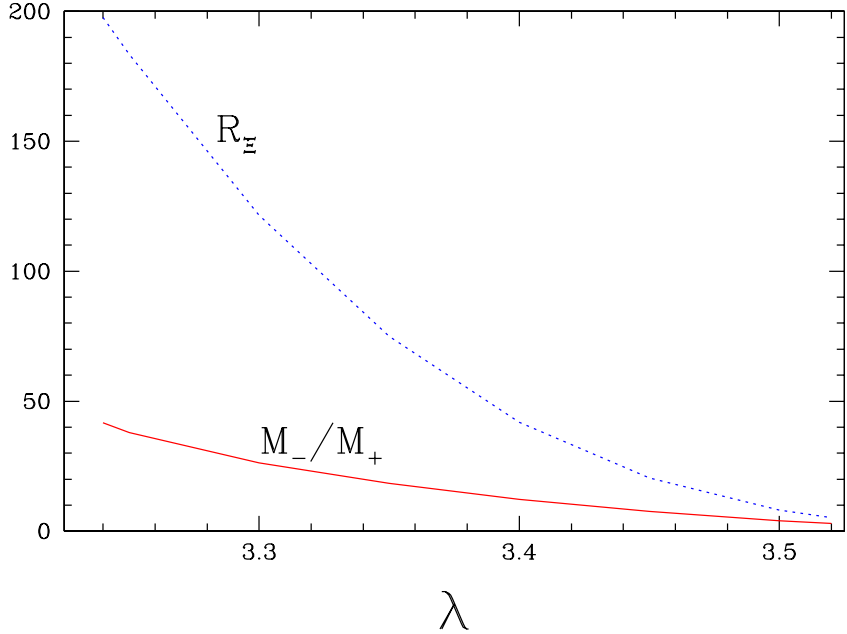}
\caption{\label{f4} \small{
The variation of the shock strength (solid line, red coloured in the online version) and
the entropy accration rate ratio at the shock (dotted line, blue coloured in the online 
version) as a function of the flow angular momentum for a fixed set of initial boundary 
conditions $\left[{\cal E}=1.0000033,\gamma=4/3,a=0.3\right]$ Shock strength anti-correlates with
the shock location. As a result, shocks formed closer to the black hole are stronger.
}}

\end{center}
\end{figure}

\section{A Lyapunov like treatment}
\label{section7}
\noindent
As we already have observed, for an extended region of the parameter 
space describing the multi-transonic accretion flow, shock does not form in spite of 
the the availability of two saddle type sonic points to the accretion flow 
to pass through. This region has been marked as ${\rm A_{\rm NS}}$, and is 
bounded by the span of the specific angular momentum $\left[\lambda_2-\lambda_3\right]$.
One might raise the question about what dictates the flow not to make a shock transition even if 
the shocked flow can have the saddle type inner sonic point available to pass through. 
In other words, given all the favourable conditions, i.e., the presence of two 
real physical sonic points, and a larger than one entropy accretion rate ratio for the 
flows passing through these sonic points (with higher entropy accretion rate for the 
flow passing thorough the outer sonic point), what prohibits the Rankine Hugoniot condition 
to be satisfied, as it gets satisfied for the region ${\rm A_{\rm S}}$? In reality, no 
information can readily be obtained from the initial boundary conditions 
$\left[{\cal E},\lambda,\gamma,a\right]$ as well as the measure of the entropy accretion 
rate (which, of course, is purely a function of $\left[{\cal E},\lambda,\gamma,a\right]$,
and hence, does not impose any additional condition) which may provide some certain 
well defined conditions predicting about the validity of the Rankine Hugoniot conditions
over certain region of the parameter space describing the multi-transonic accretion 
flow. 

Usual trends in the contemporary literature (\citep{c89,das02} and references therein)
to explain the origin and the characteristic features of the ${\rm A_{\rm NS}}$ region is
to make hypothesis that the shock may become oscillatory in that region, in absence of the 
fine tuning condition of the entropy generation at the shock (that the difference between the 
entropy accretion rate should exactly be equal to the excess amount of ${\dot {\Xi}}$ 
produced at the shock), and the region ${\rm A_{\rm NS}}$ thus has been attributed to 
explain, in analogy with the results obtained from the full numerical time dependent 
simulation of the shocked axisymmetric flow (see, 
e.g., \citep{spon-molt,okuda2} and references therein), the formation of the quasi-periodic oscillation 
of the black hole candidates. This approach, however, is fundamentally inconsistent from the mathematical 
perspective. The parameter space corresponding to the formation of the 
${\rm A_{\rm NS}}$ region (of {\it any} region of the parameter space shown in the
figure 1) is obtained from a completely stationary configuration, where the mother 
equations have been constructed as a set of explicit time independent equations. The
notion of the oscillation of the shock (or the very phrase '`oscillation' itself) is
associated with the dynamical temporal evolution of the configuration governed by a 
set of full time dependent equations. Hence, although for full time 
dependent numerical simulations, the appearance of the shock oscillation can be 
considered as a consistent physical scenario, such shock oscillation prescriptions
can {\it never} be applied to clarify any results which has been obtained by solving a set of 
explicit steady state equations. 

The above issue prompted us to put forward a theoretical condition, which, within the framework of 
the stationary configuration, will be able to explain why the entire region of the 
multi-transonic accretion parameter space will not allow a stationary shock solution, and, instead,
only a restricted part of the region will produce a shocked accretion flow. We 
propose a Lyapunov like treatment of the transonic accretion solution,
within a complete stationary framework,  to accomplish the above task. It is not quite uncommon in 
the literature to reply on a Lyapunov like treatment, within the configuration of a stationary 
set up which contains an equilibrium point (the critical point) to study the state 
transition of such systems, subjected to a small variation of the overall parameter space. 
Such a methodology have been applied to several systems, for example, to describe 
the state transition of a spin galas phase in terms of the $T=0$ critical point 
where $T$ is the absolute temperature (\citep{bray-moore}), or to study the 
thermodynamic stability of the protein like heteropolymers against the biological 
mutation process, by considering such process as a perturbation (\citep{jkb-protein}).  
In what follows, we propose a one dimensional effective Lyapunov index ${\sf L_{\sf eff}}$.
${\sf L_{\sf eff}}$ will exclusively predict which final state -- a shocked flow 
passing through both the outer and the inner sonic points, or a shock free flow passing 
only through the outer sonic point -- a multi-transonic accretion will follow, 
subjected to the specific initial boundary conditions, and the infinitesimal variation 
of the control parameter describing the secular evolution of the flow. 

\begin{proposition}
\label{prop1}
\noindent
On a two dimensional parameter space spanned by $\left[{\cal E},\lambda\right]$ as shown in the figure 1.
let us consider two transonic solutions characterized by infinitesimally different values of 
the specific flow angular momentum $\lambda_i$ and $\lambda_f$, satisfying the condition
\begin{equation}
{\sf Lt}_{{\epsilon}{\rightarrow}0}~{\vert}\lambda_i-\lambda_f{\vert}{=}{\epsilon}
\label{l1}
\end{equation}
Let $u_1(r)$ and $u_2(r)$ be the values of
the advective velocity at a certain value of the 
radial distance $r$ for two transonic solutions
characterized by
very close values of the angular momentum $\lambda_1$ and
$\lambda_2$ respectively, with ${\cal E}$ and $\gamma$ kept fixed.
Let $u_1(r-R)$ and $u_2(r-R)$ be
 the corresponding velocities for the
above-mentioned transonic flow solutions
at the radial distance $r-R$, with the obvious requirement that
$R<r$.
We then define the
effective Lyapunov exponent ${\sf L_{\sf eff}}$, so that the
following condition holds:
\begin{equation}
\frac
{u_1(r-R)-u_2(r-R)}
{u_1(r)-u_2(r)}
\sim
e^{{\sf L_{\sf eff}}R} .
\label{l2}
\end{equation}
\end{proposition}
The effective index ${\sf L_{\sf eff}}$ can also be calculated by 
varying ${\cal E}$ instead of $\lambda$.

Without any loss of generality, one can normalize the constant to appear in front of the 
$e^{{\sf L_{\sf eff}}R}$ and can define the effective Lyapunov exponent as:
\begin{equation}
{{\sf L_{\sf eff}}}=\frac{1}{R}
\left\{
\frac
{u_1(r-R)-u_2(r-R)}
{u_1(r)-u_2(r)}
\right\}
\label{l3}
\end{equation}
It is obvious from the expression of ${{\sf L_{\sf eff}}}$ that for accretion,
the effective Lyapunov index is a measure of how drastically the output 
(Mach number variation in the flow topology) will change with respect to 
an arbitrary small perturbation in the control parameter (the specific angular 
momentum $\lambda$, for our case). Hence there is no harm in assuming that 
${{\sf L_{\sf eff}}}$ does share the properties of a Lyapunov exponent in 
the context that a higher value of the modulus of ${{\sf L_{\sf eff}}}$ indicates
a greater available volume of the phase space (in connection to the 
Kaplan-Yorke conjecture \citep{kaplan-yorke}).

We now consider two different values of angular momentum infinitesimally separated 
and lying at two different side of the value $\lambda=\lambda_1$ (the value of the angular momentum
corresponding to the boundary of the {\bf O} and the {\bf A} region of the 
parameter space diagram as shown in the figure 1), as $\lambda_{\pm}$ where 
$\lambda_-=\lambda_1-\epsilon$ and $\lambda_+=\lambda_1+\epsilon$ for $\epsilon{\rightarrow}0$.
For $\lambda=\lambda_-$, one has a mono-transonic accretion only through the outer sonic point,
and $\lambda=\lambda_+$ provides the multi-critical accretion. We now calculate the 
${{\sf L_{\sf eff}}}$ for $\left[r=10r_g,R=0.2r_g,\lambda_-=2.839,\lambda_+=2.84\right]$,
where $u_1$ corresponds to the solution characterized by $\lambda_-$ and $u_2$ corresponds to 
the solution characterized by $\lambda=\lambda_2$, all  
for the initial boundary condition $\left[{\cal E}=1.0000033,\gamma=\frac{4}{3},a=0.3\right]$
for the following two different situations as described below:
\begin{enumerate}
\item For $\lambda=\lambda_+$, we fist assume that the accretion flow, even formally being multi-transonic,
passes only through the outer sonic point (which is the reality because solution characterized by $\lambda_+$ 
is a shock free solution from ${\rm A}_{\rm NS}$). We then calculate $u_2(r)$ and $u_2(r-R)$ for 
flow passing through the outer sonic point. For such situation, we denote the effective Lyapunov index 
as ${{\sf L_{\sf eff}}}^{\left[\lambda_-^{\rm out},\lambda_+^{\rm out}\right]}$,
where it is understood that the mono-transonic accretion characterized by $\lambda_-$ passes only 
through the outer sonic point and the the effective Lyapunov index is being calculated at 
$r=10r_g$ for accretion solution characterized by $\lambda_+$, which has 
been considered to pass through the outer sonic point, where the outer sonic point is larger than 
$r=10r_g$. Within the bracket, $\lambda_-^{\rm out}$ has been written first, and then 
$\lambda_+^{\rm out}$. This sequence indicates that the change of the specific angular 
momentum has been made such a was that a transition is being induced from the mono-transonic 
accretion (passing through the outer sonic point) characterized by $\lambda_-$ to the
multi-transonic accretion (but here again, passing only through the corresponding outer sonic
point, and does not encounter the inner sonic point because shock does not form in this 
case) characterized by $\lambda_+$. The modulus of the numerical value of the effective Lyapunov 
index, which has been denoted by 
${\sf mod}\left({{\sf L_{\sf eff}}}^{\left[\lambda_-^{\rm out},\lambda_+^{\rm out}\right]}\right)$,
is found to be $\sim {1.6631}$  for the present situation.
\item Here for $\lambda_+$, we assume, for a hypothetical situation, what would have been 
the value of the effective Lyapunov index at $r=10r_g$, if, for whatever reason, the muli-critical 
accretion would finally pass through the inner sonic point. Hence $u_2(r)$ and $u_2(r-R)$ have been
calculated for flow passing through the inner sonic point, and the corresponding effective 
Lyapunov index 
${{\sf L_{\sf eff}}}^{\left[\lambda_-^{\rm out},\lambda_+^{\rm in}\right]}$ has been computed. The 
the corresponding modulus of the numerical value of 
${{\sf L_{\sf eff}}}^{\left[\lambda_-^{\rm out},\lambda_+^{\rm out}\right]}$, which is 
denoted as
${\sf mod}\left({{\sf L_{\sf eff}}}^{\left[\lambda_-^{\rm out},\lambda_+^{\rm in}\right]}\right)$
is found to be ${\sim}1.60738$. 
\end{enumerate}

We thus clearly have
\begin{equation}
{\sf mod}\left({{\sf L_{\sf eff}}}^{\left[\lambda_-^{\rm out},\lambda_+^{\rm out}\right]}\right)
>
{\sf mod}\left({{\sf L_{\sf eff}}}^{\left[\lambda_-^{\rm out},\lambda_+^{\rm in}\right]}\right)
\label{l4}
\end{equation}
We now apprehend what does the above inequality indicate.

If one makes a transition from the mono-transonic flow to the multi-transonic flow by an 
{\it incremental} smooth change of the specific angular momentum, 
${{\sf L_{\sf eff}}}$ is a essentially a measure of the probability of the outcome of 
final state corresponding to such a transition. The effective Lyapunov exponent has been 
defined in such a way that the above findings allows us to make the following statement:

\begin{conjecture}
\label{conjecture1}
\noindent
For a certain transonic accretion state characterized by a control parameter $c_i$
($c_i=\lambda_-=\lambda_1-\epsilon$ for the present purpose), if there are two 
possible transitions to two different transonic states, both characterized by the same 
value of the final control parameter $c_f$ ($c_f=\lambda_+=\lambda_1+\epsilon$ for the 
present purpose), differing from $c_i$ by an infinitesimal measure
$\epsilon{\rightarrow}0$, and if such transitions can be quantified by two 
different effective Lyapunov indices 
${{\sf L^1_{\sf eff}}}$ and ${{\sf L^2_{\sf eff}}}$, then the transition characterized by the 
higher absolute value (the modulus of the numerical value) of the effective Lyapunov 
index is a more probable outcome with greater likelihood, since a higher value of the 
Lyapunov index indicates a 
greater available occupancy of the phase volume for the solution. 
\end{conjecture}

This means that according to the inequality (\ref{l4}), the probability of finding 
(at $r=10r_g$) the accretion flow passing through the outer sonic point (shock free 
multi-criticality) is higher than the probability of finding it to pass through the
inner sonic point (shocked multi-transonic accretion). And that is exactly what is 
observed for the ${\sf A_{\sf NS}}$ region. We calculate, for a wide range of $r$ and
$R$, with $r$ lying close to the inner sonic point, that is, closer to the black hole in comparison to
any `anticipated' artificial shock location, so as to calculate the ${{\sf L_{\sf eff}}}$
for an `artificial' post shock flow (had it been the case that the shock would, at all,
form, which clearly does not, in reality). This procedure {\it guarantees} the consistency 
of checking the probability of two possible outcomes at a length scale where one can distinguish 
a shocked and a shock free solution. For all possible configurations, we consistently 
found that the inequality (\ref{l4}) always holds good.

What about the {\it true} multi-transonic accretion flow structure allowing a standing shock? We can 
show that sufficient room for such flow configuration is available as well. For this purpose,
consider two infinitesimally separated value of $\lambda$, $\lambda_{\pm}$ at the opposite sides of the 
boundary separating the ${\sf A}$ and the ${\sf A_1}$ region. Here $\lambda_-=\lambda_5-\epsilon$
provides a multi-transonic accretion and $\lambda_+=\lambda_5+\epsilon$ provides a
transonic accretion passing only through the inner sonic point (and associated with a 
homoclinic orbit characterized by an additional pair of critical point, none of 
which a real physical accretion solution can pas through). We now calculate the 
corresponding effective Lyapunov indices for a 
{\it decremental} smooth change of the angular momentum. The unique initial state
(accretion passing only through the inner sonic point) is characterized by $\lambda_+$.
Two final states, both characterized by $\lambda_-$, are available. The first option for 
such final state is a hypothetical shock free multi-critical accretion passing through 
the outer sonic point only, on which $u_2(r)$ and $u_2(r-R)$ are measured at some 
specified value of $r$ and $R$. The corresponding effective Lyapunov index being 
${{\sf L_{\sf eff}}}^{\left[\lambda_+^{\rm in},\lambda_-^{\rm out}\right]}$.
For the other variant (the actual case), the shock forms and the accretion finally 
passes through the inner sonic point. The corresponding index is 
${{\sf L_{\sf eff}}}^{\left[\lambda_+^{\rm in},\lambda_-^{\rm in}\right]}$.

A suitable choice of $r$ is trivial, it has to be smaller than the smallest possible shock 
location for $\lambda_-$ type flow, so as to the length scale for our interest lies 
in the post-shock region. All possible variation of such $r$ and $R$ confirms the 
generic nature of the following inequality
\begin{equation}
{\sf mod}\left({{\sf L_{\sf eff}}}^{\left[\lambda_+^{\rm in},\lambda_-^{\rm in}\right]}\right)
>
{\sf mod}\left({{\sf L-{\sf eff}}}^{\left[\lambda_+^{\rm in},\lambda_+^{\rm out}\right]}\right)
\label{l5}
\end{equation}

\begin{lemma}
\label{lemma1}
\noindent
For an incremental smooth change of the flow specific angular momentum,
a transition from a mono-critical mono-transonic flow to a multi-critical flow 
will not produce a shock, however, for a  decremental smooth change of $\lambda$, a 
transition from a mono-transonic accretion to a multi-critical accretion 
must accompany a standing shock, and the multi-critical accretion is a 
true multi-transonic one.
\end{lemma} 

It is interesting to note that the above lemma has one to one correspondence 
with our findings related to the hysteresis effect as described in the previous section.

We thus define an independent quantity, within a framework of a purely 
stationary configuration, namely the effective Lyapunov index
${{\sf L_{\sf eff}}}$, which, for the first time in literature, can analytically 
explain why one obtains shock free solutions even for a multi-critical flow configuration. 
\section{Discussion}
\label{section8}
In the present paper we study the topology of the low angular momentum flow onto a black hole as a function 
of the flow global parameters. Such a study may indicate the behaviour of the time evolution of the flow 
seen as a sequence of quasi-stationary stages.

For a given set of other parameters (the Bernoulli constant, adiabatic index and the Kerr parameter) 
the change in the angular momentum of the flow (determined by the outer boundary conditions) leads to 
a sequence of solutions passing through two discontinuous changes in the flow topology, and consequently, 
in the flow local parameters.

The discontinuity at $\lambda_5$ corresponds to a heteroclinic solution (a `fish' topology) with a flow 
line connecting the inner and the outer critical points. As we argue in Sect.~\ref{section6}, the physical 
flow is likely to change discontinuously at $\lambda_5$ if the angular momentum increases from 
$\lambda < \lambda_5$ to $\lambda > \lambda_5$ as a result of systematic secular evolution, 
and then solutions both for $\lambda < \lambda_5$ and $\lambda < \lambda_5$ do not develop 
a shock. However, if in the course of secular evolution  the angular momentum decreases 
from $\lambda > \lambda_5$ to $\lambda < \lambda_5$ the systems evolves through 
$\lambda_5$ without showing any discontinuous change but instead develops a shock, with 
the strength of the shock increasing with the  $\lambda_5/\lambda$ ratio. 

The above mentioned finding has further been supported by a Lyapunov like treatment of the state 
transition of the flow, as discussed in detail in Sect.~\ref{section7}.

The discontinuity at $\lambda_4$ corresponds to the minimum value of the angular momentum for which 
the solutions with shocks exist. It is likely to show up if the angular momentum systematically 
decreases, first from $\lambda > \lambda_5$ to $\lambda < \lambda_5$ (following the branch with 
shocks) and then down below $\lambda < \lambda_4$. The solution must then change discontinuously 
to the one without a shock. The reverse evolution does not lead to discontinuity as the flow 
then proceeds as a shock-free solution continuously up to $\lambda_5$ where it reaches the critical 
`fish' topology.

The analytical study of the solution stability close to `fish ' topology ($\lambda \propto \lambda_5$) 
cannot be performed since infinitesimal changes in the solution parameters lead to the arbitrarily large
change in the flow parameters. The problem should be addressed through 
time-dependent dynamical simulations.

The  issue of the topological change in the flow is generic and not related to any specific 
choice of the other flow parameters. We will show in our next work (Das \& Czerny, in preparation) 
that the `fish' topology always appears as a boundary between the inflow through the inner critical 
point and through the outer critical point, irrespective of the flow geometry (as long as it contains 
some non zero angular momentum), the specific metric, and the choice of the equation of state.

Our discussion shows that both solutions with, and without a shock, are likely to be met in real 
sources accreting material with low angular momentum, and the actual choice of the solution by 
the system depends on the flow history.

If the accretion parameters in an astrophysical system are in the range somewhat broader than  
$\lambda_4$ - $\lambda_5$ and vary in time the system will show sudden rapid changes in the 
luminosity accompanying the sudden flow restucturization. We qualitatively predict the appearance 
of flares with more rapid increase of the luminosity (when crossing from $\lambda < \lambda_5$ 
to $\lambda > \lambda_5$) followed by a slow decay (evolution partially along the shock branch 
down to $\lambda < \lambda_4$). Quantitative estimates would require computations of the 
flow emissivity.   

The appearance of the homoclinic orbit on the phase portrait is not unique for the 
axisymmetric rotating flows only. For spherically symmetric accretion with self gravity 
and radiative processes incorporated into it \cite{malec1,malec2}, 
such homoclinic paths would also appear \cite{thesis}, 
making the spherically symmetric accretion to be a bi-critical flow. It appears that 
the emergence of a homoclinic orbit is due to the smooth decrement of the 
radial inflow velocity, due to the centrifugally supported barrier for rotating flow, 
while due to the radiative losses for the spherical accretion.
 
In this work, we have used a relatively lower value of the Kerr parameter ($a=0.3$) as a 
representative one. This does not, by any means, limit the applicability of our 
methodology for the case of black holes possessing higher value of the spin angular 
momentum because the Kerr parameter has been treated as a free parameter in our 
construction and any value of $a$ can easily be plugged in into our model set up.
For example, we obtain consistent shock location for the value of $a$ as high as $0.99$ 
(ad for values of $\gamma$ other than 4/3) as well. We can, compute the shock location and 
all other shock related quantities in general as a function of the Kerr parameter as well, and 
that has been presented in some other work (Barai et al. in preparation). 

Some of the results presented in this paper may also
apply to GRB. In this event, modeled as a failed supernova,
the star first collapses due to the fuel exhaustion, the
collapse is halted close to the core (at $\sim 500 $km)
and replaced by expansion. Part of the stellar envelope
is rejected but the neutrino cooling stops part of the
material and subsequently the  slow fallback stage takes place
\cite{mwh}
onto a newly formed compact object.
This accretion stage of the long gamma-ray burst
can be described within the frame of our model assuming
that the
angular momentum of the back falling envelope is not large.
If the accretion flow proceeds through a solution with a
shock, the shock-related periodic oscillations may in a
natural way explain for example the fast optical variability
of the Naked-Eye burst GRB080319B \cite{beskin}.


\section*{Acknowledgments}
TKD would like to acknowledge the warm hospitality provided by the 
Polish Academy of Sciences, through CAMK, Warsaw, Poland, in the
form of a visiting scientist. This work is partially supported by
the grants N203 011 32 1518 and N N203 38 0136, 
and by the Polish Astroparticle Network 621/E-78/BWSN-0068/2008.
Stimulating discussions with Jayanta Bhattacharjee
and Micha\l~ R\' o\. zyczka are acknowledged.

\end{document}